\documentclass[%
 reprint, 
 superscriptaddress,
 amsmath,amssymb,
 aps, physrev,
floatfix,
]{revtex4-2}

\usepackage{graphicx}
\usepackage{dcolumn}
\usepackage{bm}
\usepackage{placeins}
\usepackage{xcolor}

\begin{document}

\title{\textbf{Topological Signatures of AGN Feedback in the {\sc Simba} Simulations}}

\author{Ryusei R. Kano}
\homepage{https://kano-ryusei.net/}
\email{ryuseikano@nagoya-u.jp}
\affiliation{Division of Particle and Astrophysical Science, Nagoya University, Furo-cho, Chikusa-ku, Nagoya 464-8602, Japan}%
\affiliation{Institute for Astronomy, University of Edinburgh, Royal Observatory, Edinburgh, EH9 3HJ, UK}%

\author{Romeel Dav\'e}
\affiliation{Institute for Astronomy, University of Edinburgh, Royal Observatory, Edinburgh, EH9 3HJ, UK}%
\affiliation{Department of Physics and Astronomy, University of the Western Cape, Bellville, Cape Town 7535, South Africa}%

\author{Tsutomu T. Takeuchi}%
\affiliation{Division of Particle and Astrophysical Science, Nagoya University, Furo-cho, Chikusa-ku, Nagoya 464-8602, Japan}%
\affiliation{The Research Center for Statistical Machine Learning, the Institute of Statistical Mathematics, 10--3 Midori-cho, Tachikawa, Tokyo 190--8562, Japan}%

\author{Hayato Shimabukuro}
\affiliation{South-Western Institute for Astronomy Research, Key Laboratory of Survey Science of Yunnan Province,
Yunnan University, Kunming, Yunnan 650500, People’s Republic of China}
\affiliation{Division of Particle and Astrophysical Science, Nagoya University, Furo-cho, Chikusa-ku, Nagoya 464-8602, Japan}%

\date{\today}

\begin{abstract}
Active galactic nuclei (AGN) feedback can alter  the abundances and the spatial arrangement of galaxies within large-scale structure. We apply topological data analysis (TDA) to {\sc Simba} hydrodynamical simulations with varying AGN feedback at $z=0$ to determine which galaxy populations, environments, and spatial scales are most impacted by AGN. We first show that the galaxy two-point correlation function (TPCF) is sensitive to feedback on smaller scales $r<1\,h^{-1}\mathrm{Mpc}$. 
From TDA, we use Betti curves and persistence diagrams to quantify the robustness of connectivity and loops as a function of scale. 
Betti curves vary only weakly with feedback, but persistence diagrams show clearer differences, particularly for satellite populations. We further consider the 2-Wasserstein distance,
showing that for satellites, the loop-like persistence diagrams exceed the reference; corresponding Betti curves localize this to a small-scale dependence at $\alpha\sim0.1$--$0.4\,h^{-1}\mathrm{Mpc}$, sensitive to every feedback channel, and weaker but significant differences at $\alpha\sim1$--$3\,h^{-1}\mathrm{Mpc}$. This shows that persistence diagrams and associated Betti curves can quantify differences on different scales than the TPCF. The differences are strongest when isolating satellites in quenched halos above $M_h\gtrsim10^{12}M_\odot$. 
We find that the shifts are driven by the birth and death scales of loop-like features, rather than by (dis)appearance of persistent loop features, suggesting the feedback-driven topological differences owe primarily to quenched galaxies in massive halos. Application to large-scale redshift surveys could provide a unique test of AGN feedback-driven quenching models. 
\end{abstract}

\maketitle


\section{\label{sec:intro}INTRODUCTION}

Forthcoming surveys such as Euclid, LSST, and DESI will push measurements of large-scale structure into a regime where baryonic systematics become one of the limiting uncertainties in cosmological inference \cite{laureijs2011euclid,amendola2018cosmology,ivezic2019lsst,aghamousa2016desi}. Hydrodynamical simulations demonstrate that feedback processes, particularly from active galactic nuclei (AGN), can suppress the matter power spectrum by up to $20$--$30\%$ at non-linear scales ($k \sim 10\,h\,\mathrm{Mpc}^{-1}$) \cite{van2011effects,salcido2023sp,gebhardt2024cosmological}. A challenge is that not only does feedback change galaxy abundances, but that different subgrid feedback prescriptions can leave highly degenerate signatures in standard pair statistics and higher-order moments \cite{arico2021simultaneous,nicola2022breaking,saha2024quantifying}. In this regime, a statistic that characterizes the field by a single clustering amplitude or a small set of pairwise moments can remain nearly degenerate even when the underlying structures have been reorganized in a physically meaningful way.  It is therefore worth investigating approaches that more directly quantify the Cosmic Web topology traced by galaxies, to investigate whether they can provide complementary constraints on AGN feedback.

Topological data analysis (TDA), with persistent homology as one of its central tools, is well suited to quantifying complex topologies because it tracks how Web-like features such as connected components, loops, and voids appear and disappear as the filtration scale is varied. TDA has been successfully applied to constrain primordial non-Gaussianity, estimate neutrino masses, and classify the cosmic web. Previous cosmological applications have already shown that Betti numbers and Betti curves can be useful for parameter constraints and void-related morphology \cite{van2011alpha,pranav2017topology,biagetti2021persistence,biagetti2022fisher,jalali2024imprint}. More recently, persistence diagrams have also been applied to one-dimensional HI-21 cm forest spectra, showing that the hierarchical birth--merger history of absorption troughs can separate X-ray heating from dark-matter free-streaming effects even when standard amplitude-based statistics remain strongly degenerate \cite{shimabukuro2026topological}.

However, for three-dimensional large-scale structure, the majority of TDA studies to date rely on count-based projections such as Betti numbers and Betti curves \cite{van2011alpha,pranav2017topology,biagetti2021persistence,biagetti2022fisher,jalali2024imprint,wilding2021persistent}. 
At a fixed filtration scale, a Betti number records the number of generators alive at a given scale but does not record the individual $(b_i,d_i)$ birth/death scales of those generators. For instance, the first Betti number $\beta_1$ denotes the number of loop-like one-dimensional features present at a given filtration (i.e. smoothing) scale. 
A deficit in $\beta_1$ can arise because a loop is destroyed, because its characteristic scale has shifted, or because several features have been redistributed in birth--death space in a way that leaves only a small net change in the total count \cite{wilding2021persistent,biagetti2021persistence}. In other words, Betti numbers tell us that topology changes, but not how that change occurred. 

Persistence diagrams retain that missing geometric information. To robustly quantify differences between these diagrams beyond simple counting, we turn to optimal transport theory---and specifically the Wasserstein distance. Optimal transport has a longer history in cosmology \cite{frisch2002reconstruction,takeuchi2026wasserstein}, while Wasserstein distances between persistence diagrams have more recently been applied to cosmic-web topology \cite{tsizh2023wasserstein}. In a persistence diagram, the diagonal $b=d$ represents zero-persistence topologically-trivial features that only appear at one specific scale, and a finite-persistence generator has $b<d$. The 2-Wasserstein distance evaluates the optimal matching cost between two persistence diagrams, in which each generator is matched either to a generator in the other diagram or to the diagonal: matching to the diagonal measures the cost of treating a generator as effectively removed, while matching to a generator in the other diagram measures a shift in that feature's birth and death scales. This optimal-transport viewpoint makes TDA a diagram-level diagnostic of structure formation rather than only a count-based summary like Betti numbers.

AGN feedback may alter where quenched satellites reside within halos, around halo outskirts, and along connections between neighboring halos. Such differences need not produce a large change in the two-point correlation function, but they can shift the characteristic scales of loop-like configurations traced by galaxies. Recent simulations and observations indicate that AGN feedback can influence satellite galaxies anisotropically and with a radial dependence extending to several virial radii \cite{dashyan2019agn,eckert2021feedback,goubert2024role}. Galactic conformity and anisotropic satellite quenching \cite{henriques2017galaxy,kawinwanichakij2016satellite,karp2023anisotropic} therefore motivate a search for feedback-dependent spatial differences in the quenched population. We refer to galaxy configurations that contribute to these loop-like features as ``loop-forming configurations''; this is a topological description rather than a distinct physical network. 
Persistent homology provides a way to test for such differences through changes in the birth, death, and persistence scales of the associated loop-like features. Connectivity and enclosed voids are captured in an analogous way. We call feedback-dependent differences in the locations of quenched galaxies ``spatial quenching bias.''  Our goal is thus to quantify spatial quenching bias and how it varies with AGN feedback.

In this paper, we use the feedback variants in the {\sc Simba} simulation suite \cite{dave2019simba,scharre2024effects} to explore feedback-dependent spatial differences under various assumptions of feedback, and analyze whether they are concentrated in particular galaxy populations, host-halo masses, and physical scale ranges. We compute the galaxy two-point correlation function and Betti curves to provide conventional and count-based comparisons, and contrast them with persistence diagrams and the 2-Wasserstein distance ($W_2$) which measure changes in the characteristic scales and persistence of individual topological features. To avoid mistaking changes in overall sampling density for a physical differences, we perform each topological comparison at fixed tracer number density, corresponding to fixed galaxy number density in each simulation.

More specifically, we seek to answer four questions: (1) What feedback-dependent differences are visible in the galaxy two-point correlation function, and what additional population- and scale-resolved information is provided by persistence diagrams? (2) Do the models primarily differ in the number of loop-like galaxy configurations, or in the spatial scales over which those configurations appear? In persistence-diagram terminology, these possibilities correspond predominantly to diagonal and off-diagonal transport, respectively. (3) Are the differences concentrated in satellite galaxies or in particular host-halo-mass regimes? (4) Are they more clearly associated with quenched galaxies than with the star-forming population, after controlling for galaxy number density?


in \S\ref{sec:methodology} we describe the simulations and the topological analysis methodology.  In \S\ref{sec:results1} we compute the two-point correlation function and Betti numbers for our different models.  In \S\ref{sec:results2} we focus on the differences seen in the satellite galaxy population, which shows more sensitivity to AGN feedback.  In \S\ref{sec:results3} we break down the samples in terms of halo mass and star formation rate to see trends in spatial quenching bias.  \S\ref{sec:discussion} we discuss our results to synthesize an overall scenario for how feedback induces spatial quenching bias.  In \S\ref{sec:summary} we summarize our results.

\section{METHODOLOGY}\label{sec:methodology}

\subsection{{\sc Simba} Simulations and Controlled Galaxy Catalogs}\label{sec:catalog_construction}

The {\sc Simba} simulations model a randomly-selected region of the Universe assuming a concordance $\Lambda$ Cold Dark Matter cosmology with $\Omega_{\rm matter}=0.3$, $\Omega_{\rm \Lambda}=0.7$,
and a present-day expansion rate of $H_0=70$~km~s$^{-1}$~Mpc$^{-1}$, from a redshift $z=249\to 0$ using a highly modified version of the GIZMO hydrodynamics code.  Full details are presented in \cite{dave2019simba}.

We analyze the {\sc Simba}-50 suite with identical initial conditions across four feedback realizations: Full Physics, No-Xray, No-Jet, and No-AGN \cite{dave2019simba,scharre2024effects}.  Starting from the Full Physics run, No-Xray removes X-ray AGN feedback, No-jet additionally removes kinetic bipolar AGN jet feedback, and No-AGN additionally removes radiative wind AGN feedback (resulting in all AGN feedback being off).  Note that only the Full Physics model is tuned to match observations; the others explicitly do not match observations such as galaxy stellar mass function or galaxy quenched fractions, and purely serve as numerical experiments to isolate the impact of an individual AGN feedback module.  All results in this paper use the $z=0$ snapshot of a periodic comoving cubic box with side length $L_{\mathrm{box}}=50\,h^{-1}\mathrm{Mpc}$.
Here, ``matched phases'' means that the four realizations start from the same initial density-field realization (the same random seed, Fourier-mode phases, and number of particles) and differ only in the feedback modules applied during the hydrodynamical evolution.
This mitigates the cosmic-variance contribution to the model-to-model comparisons, but it represents only a change due to the physics variations, not the cosmological representativeness of the underlying box. 

Galaxy, halo, and subgroup properties are taken from the publicly-available CAESAR catalog generated from the $z=0$ simulation snapshots. Halo masses are defined as $M_{200c}$. Within each halo, CAESAR defines the galaxy with the largest stellar mass as the central and all other member galaxies as satellites. For the star-forming/quenched split, we use the specific star-formation rate and classify galaxies with $\mathrm{sSFR} > 10^{-11}\,\mathrm{yr}^{-1}$ as star-forming and those with $\mathrm{sSFR} \le 10^{-11}\,\mathrm{yr}^{-1}$ as quenched.

Persistent homology is sensitive to the sampling density of the input point cloud. If one feedback realization contains fewer tracers (i.e. galaxies) than another, the typical inter-point spacing increases and the birth and death scales in the persistence diagrams can shift systematically to larger $\alpha$ even if the underlying geometric pattern were otherwise unchanged. Unequal tracer abundances could therefore mimic a feedback-induced topological signal. 
In {\sc Simba}, galactic feedback suppresses star formation across all masses, reducing the abundance of galaxies at both the low- and high-mass ends of the stellar-mass function \cite{dave2019simba,appleby2020impact}.
To account for this, we construct our catalogs for topological analysis at a fixed tracer number density among all models. We apply this fixed-density construction separately to every comparison sample used in this paper, including the full galaxy catalog, halo centers, centrals, satellites, galaxies in each fixed host-halo-mass bin, and the star-forming and quenched subsamples, rather than only to the full catalog. Operationally, for each galaxy comparison sample $S$, we rank the objects in that sample by stellar mass within each of the four feedback realizations, and keep only the top
\begin{equation}
N_{S,\mathrm{target}} = \min_{r\in\mathcal{R}} N_S^{(r)}
\end{equation}
of them, where $\mathcal{R}=\{\mathrm{Full\ Physics},\mathrm{No\mbox{-}Xray},\mathrm{No\mbox{-}Jet},\mathrm{No\mbox{-}AGN}\}$ and $N_S^{(r)}$ is the number of resolved objects in sample $S$ for realization $r$: that is, $N_{S,\mathrm{target}}$ is simply the smallest of the four counts for that particular sample. For the halo-center catalog, which has no stellar-mass selection, $N_S^{(r)}$ instead counts halo tracers rather than stellar-mass-ranked galaxies; in practice the halo count already matches across all four feedback realizations ($N_S^{(r)}=57927$ for every $r$), so no additional rank-based truncation is applied there. Every sample is therefore homogenized against its own counterparts, not against the total-catalog count. For example, satellites are matched to whichever of the four realizations has the fewest satellites, and the star-forming and quenched subsamples are each matched separately. For the full galaxy catalog specifically, Full Physics happens to have the fewest galaxies, because it has the strongest feedback, so it sets $N_{\mathrm{target}}$ in that case; for every other sample, whichever realization is least populated in that particular sample plays the same role. We note that since the stellar mass and halo mass are tightly correlated, this approximately preserves halo mass rank ordering, but we choose stellar mass in order to make it more connected to an observable quantity.
The resulting target numbers and median stellar and host-halo masses of the selected catalogs are summarized in Table~\ref{tab:sample_summary}.

\begin{table*}
\caption{
Summary of the samples used in the topological analyses.
$N_{\rm target}$ is the matched tracer number used in each fixed-density comparison.
The listed stellar and halo masses are medians of the final selected catalogs.
For rank-selected galaxy samples, the effective stellar-mass threshold can vary among feedback realizations because the fixed-density construction matches the tracer number rather than imposing a common absolute stellar-mass cut.
For the halo-center catalog, stellar masses are computed from the nearest associated galaxy and are shown only as a descriptive reference.
}
\label{tab:sample_summary}
\resizebox{\textwidth}{!}{%
\begin{ruledtabular}
\begin{tabular}{llccc}
Sample & Selection & $N_{\rm target}$
& median $\log_{10}(M_\star/M_\odot)$
& median $\log_{10}(M_{200c}/M_\odot)$ \\
\hline
Total galaxies
& All CAESAR galaxies; min $\log_{10}(M_\star/M_\odot)=8.42$
& 6922 & 9.34 & 11.57 \\

Halo centers
& Halo positions; no stellar-mass selection
& 57927 & 9.35 & 10.16 \\

Centrals
& Central galaxies
& 4500 & 9.41 & 11.28 \\

Satellites
& Satellite galaxies
& 2422 & 9.22 & 12.86 \\

\hline
Host-mass bin 1
& $11.5 \leq \log_{10}(M_{200c}/M_\odot) < 12.0$
& 1416 & 9.76 & 11.57 \\

Host-mass bin 2
& $12.0 \leq \log_{10}(M_{200c}/M_\odot) < 12.5$
& 790 & 9.96 & 12.04 \\

Host-mass bin 3
& $12.5 \leq \log_{10}(M_{200c}/M_\odot) < 13.0$
& 569 & 9.65 & 12.52 \\

Host-mass bin 4
& $13.0 \leq \log_{10}(M_{200c}/M_\odot) < 13.5$
& 474 & 9.50 & 12.99 \\

\hline
Quenched galaxies
& ${\rm sSFR} \leq 10^{-11}\,{\rm yr}^{-1}$
& 1457 & 10.04 & 12.55 \\

Star-forming galaxies
& ${\rm sSFR} > 10^{-11}\,{\rm yr}^{-1}$
& 4863 & 9.20 & 11.33 \\

Quenched satellites
& Satellite galaxies with ${\rm sSFR} \leq 10^{-11}\,{\rm yr}^{-1}$
& 1224 & 9.45 & 13.44 \\

Star-forming satellites
& Satellite galaxies with ${\rm sSFR} > 10^{-11}\,{\rm yr}^{-1}$
& 1170 & 9.15 & 12.29 \\
\end{tabular}
\end{ruledtabular}%
}
\end{table*}

We will contrast our topological results to a more conventional approach, namely 
the real-space galaxy two-point correlation function (2PCF) from the CAESAR galaxy catalogs, before the fixed-number-density selection used for the topological comparisons. We use the natural estimator $\xi(r)=DD(r)/RR(r)-1$, with the random-pair expectation $RR$ evaluated analytically from the spherical-shell volume accounting for periodic boundaries.

\subsection{Persistence Diagrams, Betti Curves, and Alpha Filtration}
From each galaxy catalog we filter at various scales $\alpha$ and compute persistent homology in dimensions $H_0$, $H_1$, and $H_2$ \cite{edelsbrunner1994three,edelsbrunner2002topological,zomorodian2004computing, edelsbrunner2010computational}. 
We first construct Voronoi and Delaunay tesselations from the input point set $X=\{x_i,y_i,z_i\}$, and build the filtration from Delaunay simplices whose associated Voronoi regions intersect a sphere of radius $\alpha$ around the input points.
We adopt the $\alpha$-complex because its Voronoi--Delaunay construction has a natural geometric interpretation for three-dimensional point distributions and has been used in previous topological studies to suppress spurious over-connections in dense regions while extracting the local skeleton of the cosmic web \cite{van2011alpha}. Formally, for Voronoi cells $V_i$ and Delaunay triangulation $\mathrm{Del}(X)$, the associated $\alpha$-complex may be written as
\begin{equation}
C_{\alpha}(X)
=
\left\{\sigma\in \mathrm{Del}(X)
\;\middle|\;
\bigcap_{x_i\in\sigma}\left[B_{\alpha}(x_i)\cap V_i\right]\neq\varnothing
\right\},
\end{equation}
where $B_{\alpha}(x_i)$ is the closed ball of radius $\alpha$ centered on $x_i$. We then compute the homology groups $H_k(C_{\alpha})$, whose physical interpretations are connected components ($k=0$), loops ($k=1$), and enclosed voids ($k=2$), and define the Betti numbers as their ranks,
\begin{equation}
\beta_k(\alpha) = \mathrm{rank}\,H_k(C_{\alpha}) = \dim H_k(C_{\alpha}).
\end{equation}
Each persistent generator $i$ in homology dimension $k$ appears at filtration scale $\alpha=b_i$ and disappears at $\alpha=d_i$, yielding a point $(b_i,d_i)\in D_k$ in the persistence diagram. The corresponding Betti curve is recovered from the same diagram as
\begin{equation}
\beta_k(\alpha) = \sum_{i\in D_k} \mathbf{1}\!\left(b_i \le \alpha < d_i\right),
\end{equation}
that is, as the number of $k$-dimensional generators alive at scale $\alpha$. Whenever Betti curves are shown with shaded error envelopes, those envelopes are obtained by recomputing the curves over the same repeated $80\%$ subsamples used later for the Wasserstein analysis.

To test whether a feedback-dependent difference between two Betti curves $\beta_k^A(\alpha)$ and $\beta_k^B(\alpha)$, restricted to a filtration-scale window $W$, exceeds the level expected from finite sampling alone, we use two complementary functional statistics. The first is a window-restricted form of the Kolmogorov--Smirnov-type Betti-curve distance $D_{\rm KS}$ (and its finite-sample version $D_q$) defined by \cite{chung2019exact} immediately before their Theorem 3: the maximum absolute separation within $W$,
\begin{equation}
D_{\max}(W) = \max_{\alpha\in W} \left| \beta_k^A(\alpha) - \beta_k^B(\alpha) \right|,
\end{equation}
The second is motivated by the integrated functional distance in Eq.~(15) of \cite{cisewski2022differentiating}, which we adapt to a squared ($L^2$-type) integrand evaluated in $\ln\alpha$ and restricted to the window $W$,
\begin{equation}
T_{\rm area}(W) = \int_W \left[\beta_k^A(\alpha) - \beta_k^B(\alpha)\right]^2 \, d\ln\alpha.
\end{equation}
Related integrated representations of the Betti function, obtained by averaging it over sub-intervals of the filtration range, also underlie the permutation-test statistics of \cite{islambekov2024vector} (their Eqs.~3, 4, and 9), though $T_{\rm area}$ itself is not one of their defined statistics. $D_{\max}$ is sensitive to a localized deviation, while $T_{\rm area}$ is sensitive to a broad, coherent separation spread across the window.

We assess significance by comparing an observed inter-model statistic with the distribution of the same statistic evaluated between subsamples of the same Full Physics realization, following the tail-counting form of the permutation $p$-value in Eq.~(16) of \cite{cisewski2022differentiating} and the $(Z+1)/(N+1)$ finite-sample correction established in Theorem~1 and Algorithm~2 of \cite{robinson2017hypothesis}. Our null construction differs from both of these in one respect: rather than permuting diagram labels between two groups, we build the null from same-realization subsample pairs, since our comparison is between galaxy catalogs rather than between two exchangeable labeled groups. It also differs from \cite{chung2019exact}, who derive the null distribution of $D_q$ combinatorially from a monotonic graph filtration; our $\alpha$-complex filtration admits no analogous closed-form null, so we calibrate both statistics empirically instead. We draw $B=5000$ independent pairs of freshly resampled $80\%$ Full Physics subsamples; for each pair, we recompute persistent homology and the corresponding Betti curve for each of the two subsamples independently, and evaluate each statistic $S\in\{D_{\max},T_{\rm area}\}$ on that one pair, giving $B=5000$ mutually independent same-realization reference values $\{S^{\rm null}_b\}_{b=1}^{B}$; no subsample is reused across pairs. These values give the empirical null distribution expected from finite-sampling variation alone. For an observed inter-model value $S_{\rm obs}$, we estimate the empirical $p$-value as
\begin{equation}
p = \frac{1+\sum_{b=1}^{B} \mathbf{1}\!\left(S^{\rm null}_b \ge S_{\rm obs}\right)}{B+1},
\end{equation}
where $\mathbf{1}(\cdot)$ is the indicator function.

Although we compute persistent homology up to dimension $H_2$, the primary transport analysis and controlled subpopulation diagnostics focus on $H_0$ (connectivity) and $H_1$ (loops). We treat $H_2$ void statistics as descriptive only, because robust void measurements in the $50\,h^{-1}\mathrm{Mpc}$ {\sc Simba}-50 volume are especially susceptible to finite-volume effects. The relevant scale separation is physical: cosmological void diameters are commonly of order $10$--$50\,h^{-1}\mathrm{Mpc}$, comparable to the side length of the box used here, so $H_2$ generators can be sensitive to the finite volume and to the way large cavities interact with the boundaries of the fundamental simulation domain \cite{pan2012cosmic,sutter2012public,wilding2021persistent}. By contrast, the $H_0$ and $H_1$ response targeted in this work is associated primarily with intra-halo, circumgalactic, and filamentary galaxy configurations on smaller scales of a few $h^{-1}\mathrm{Mpc}$, which are sampled many times within the volume.

\subsection{Wasserstein Distance and Optimal Matching}\label{sec:wasserstein_matching}
Having defined the persistence diagrams $D_n$, we now quantify differences between feedback models by endowing diagram space with the standard 2-Wasserstein metric \cite{mileyko2011probability,turner2014frechet}. For two diagrams $D_n^{A}$ and $D_n^{B}$ in homology dimension $n$, we compute
\begin{equation}
W_2\left(D_n^{A},D_n^{B}\right)
=
\left(
\inf_{\gamma}
\sum_{x\in D_n^{A}}
\left\|x-\gamma(x)\right\|_{2}^{2}
\right)^{1/2},
\end{equation}
where $\gamma$ is a bijection that matches diagram points either to points in the other diagram or to the diagonal. We adopt the standard 2-Wasserstein metric (\texttt{order=2} and \texttt{internal\_p=2} in the GUDHI implementation) because, unlike the Bottleneck distance \cite{cohen2005stability}, it accumulates contributions from all matched generators rather than only the single largest shift. We adopt \texttt{order=2} because it gives relatively greater weight to large displacements in birth-death space than \texttt{order=1}, while remaining sensitive to contributions from all matched generators.


We assess statistical robustness with a repeated $80\%$ random subsampling procedure. For each comparison between models $A$ and $B$, we perform $N_{\mathrm{iter}}=50$ iterations in which we draw random $80\%$ subsamples from each catalog and recompute the persistence diagrams and their $W_2$ distance. Writing the resulting distances as $\{W_{2,n}^{A,B,(j)}\}_{j=1}^{N_{\mathrm{iter}}}$ for homology dimension $n$, we summarize the inter-model signal by its median,
\begin{equation}
\widetilde{W}_{2,n}^{A,B} = \mathrm{median}\!\left\{W_{2,n}^{A,B,(j)}\right\}.
\end{equation}
To construct the reference Full Physics value used in the main figures, we repeat the same procedure using two separately drawn (overlapping) $80\%$ subsamples from the Full Physics realization, producing an ensemble $\{W_{2,n}^{\mathrm{same},(j)}\}$. We then define the empirical reference band and the 84th-percentile finite-sampling reference by
\begin{equation}
\begin{aligned}
\mathcal{N}_{2,n}
&= \left[
Q_{0.16}\!\left(W_{2,n}^{\mathrm{same}}\right),
\right.\\
&\qquad\left.
Q_{0.84}\!\left(W_{2,n}^{\mathrm{same}}\right)
\right],\\
W_{2,n}^{\mathrm{limit}}
&= Q_{0.84}\!\left(W_{2,n}^{\mathrm{same}}\right).
\end{aligned}
\end{equation}
where $Q_p$ denotes the empirical $p$th quantile of the same-run ensemble. For visual consistency, the main transport figures use the reference band constructed from the Full Physics realization. This choice is used only as a common finite-sampling reference within the fixed {\sc Simba}-50 volume, not as a formal hypothesis test or as an externally calibrated significance threshold. To test whether our conclusions depend on this reference choice, Appendix~\ref{app:reference_sensitivity} repeats the same-run construction using each of the four feedback realizations as the reference catalog; we find no substantive change in our conclusions. We report the result as a reference-realization sensitivity test, asking whether each adjacent feedback-step median exceeds the 84th percentile of the same-run reference distribution constructed from Full Physics, No-Xray, No-Jet, or No-AGN. In the transport figures, the bars report $\widetilde{W}_{2,n}^{A,B}$, the error bars span the 16th to 84th percentiles of $\{W_{2,n}^{A,B,(j)}\}$, and the gray shaded band marks $\mathcal{N}_{2,n}$.  In order to identify a feedback-dependent difference that exceeds the same-run finite-sampling reference, we require $\widetilde{W}_{2,n}^{A,B} > W_{2,n}^{\mathrm{limit}}$, i.e. that the difference exceeds the 84th percentile range from the Full Physics resamplings.  $W_2$ thus serves as a measure of the difference between persistence diagrams among runs, though one should note that it is an aggregate statistic and should not be interpreted as a real-space scale shift in topological features between different models.

To track how the topology changes as feedback modules are added among the {\sc Simba} feedback variants, we define the X-ray-heating, kinetic-jet, and radiative-wind contributions in homology dimension $n$ as
\begin{equation}\label{eq:diff_homology}
\begin{aligned}
\Delta_{\mathrm{X\mbox{-}ray},n}
&= \widetilde{W}_{2,n}^{\mathrm{No\mbox{-}Xray},\mathrm{Full\ Physics}},\\
\Delta_{\mathrm{jet},n}
&= \widetilde{W}_{2,n}^{\mathrm{No\mbox{-}Jet},\mathrm{No\mbox{-}Xray}},\\
\Delta_{\mathrm{wind},n}
&= \widetilde{W}_{2,n}^{\mathrm{No\mbox{-}AGN},\mathrm{No\mbox{-}Jet}}.
\end{aligned}
\end{equation}
Each difference measures the conditional topological contrast associated with adding a given AGN feedback module within the {\sc Simba} variants hierarchy. The resulting quantities therefore identify which feedback step carries the largest transport cost for a given tracer sample, while retaining any impact of the previously active feedback modes.

AI-assisted coding tools, including Anthropic Claude Code (Claude Sonnet 5), were used during the development, debugging, and verification of the analysis pipeline used to compute the persistence diagrams, Betti curves, and Wasserstein distances reported in this work, and during the design and execution of the independent numerical checks used to verify the reported results against the underlying cached outputs. All AI-generated or AI-modified code was directed interactively by the authors, reviewed line by line, and tested against the cached intermediate results before being used to produce the results reported here.

\section{STANDARD CLUSTERING AND COUNT-BASED TOPOLOGICAL SUMMARIES}\label{sec:results1}

\subsection{Pair Statistics and Absolute Betti Curves}

\begin{figure}[!htbp]
\centering
\includegraphics[width=\columnwidth]{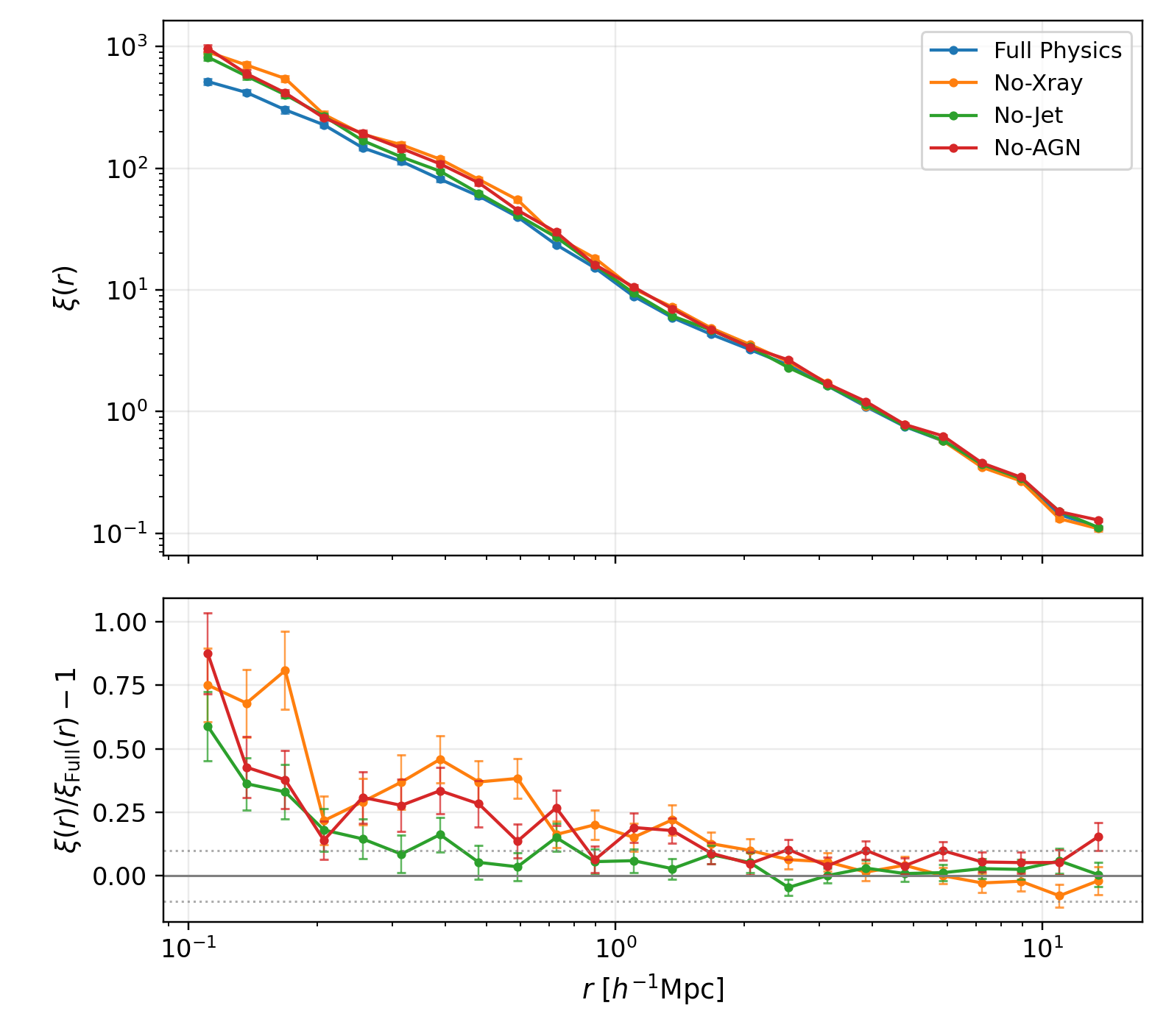}
\caption{\label{fig:2pcf} Real-space galaxy two-point correlation function $\xi(r)$ vs. comoving separation $r$, measured from the untrimmed CAESAR galaxy catalogs for each of the four {\sc Simba} feedback realizations (see legend). Error bars show the standard deviation over 50 random $80\%$ subsamples and quantify finite-sampling variation within the fixed simulation volume; they do not include cosmic variance. Differences among the models are largest on sub-Mpc scales, while the curves converge to substantially smaller relative differences beyond a few $h^{-1}\mathrm{Mpc}$.}
\end{figure}

We begin by examining a conventional two-point clustering diagnostic to see how it varies among feedback models. Figure~\ref{fig:2pcf} shows the galaxy two-point correlation function (2PCF) of the untrimmed total-galaxy catalogs. Separations are divided into 24 logarithmic bins over $0.1\le r/(h^{-1}\mathrm{Mpc})\le15$. Feedback-dependent clustering differences are largest below approximately $1$--$2\,h^{-1}\mathrm{Mpc}$, where the weaker-feedback realizations generally have a higher correlation amplitude than Full Physics. The curves converge on larger scales: beyond a few $h^{-1}\mathrm{Mpc}$, most relative differences are at approximately the $10\%$ level or below. The largest difference occurs when removing X-ray feedback, showing that this mode is primarily responsible for reducing small-scale clustering, while removing jet and then radiative AGN feedback yields results closer to the Full Physics run.  Overall, the 2PCF shows some differences, but they tend to be confined to the so-called one-halo term at sub-Mpc scales, while on larger scales there is little sensitivity to feedback.

\begin{figure*}[t]
\centering
\includegraphics[width=\textwidth]{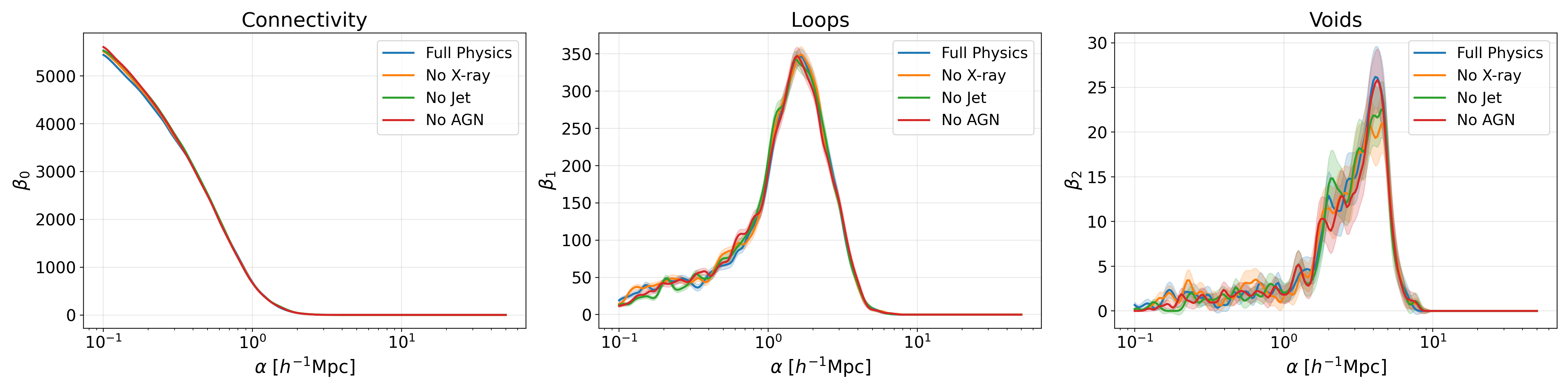}
\caption{\label{fig:global_topology} Betti numbers $\beta_0$, $\beta_1$, and $\beta_2$ (left to right panels) vs. filtration scale $\alpha$, for the total galaxy sample in each of the four {\sc Simba} feedback realizations. Shaded envelopes are derived from the repeated $80\%$ subsampling procedure used for the Wasserstein analysis. Absolute Betti curves for the total galaxy sample remain broadly stable across the feedback models, showing no large difference in the volume-averaged galaxy distribution.}
\end{figure*}

Figure~\ref{fig:global_topology} shows the Betti numbers $\beta_0$, $\beta_1$, and $\beta_2$ (left to right panels) as a function of filtration scale $\alpha$, for the total galaxy sample in each of the four feedback realizations. At the level of the full cosmic volume, the absolute Betti curves remain broadly consistent across feedback models: $\beta_0$, $\beta_1$, and $\beta_2$ tracing large-scale merger, loop, and void structures vary only weakly across every realization. Hence TDA count-based summaries show no large difference in the volume-averaged galaxy distribution, at either small or large scales.  This motivates going to more sophisticated topological analysis using the 2-Wasserstein distance as outlined previously.

\subsection{Full Galaxy and Halo Sample Results}

\begin{figure}[!htbp]
\centering
\includegraphics[width=\columnwidth]{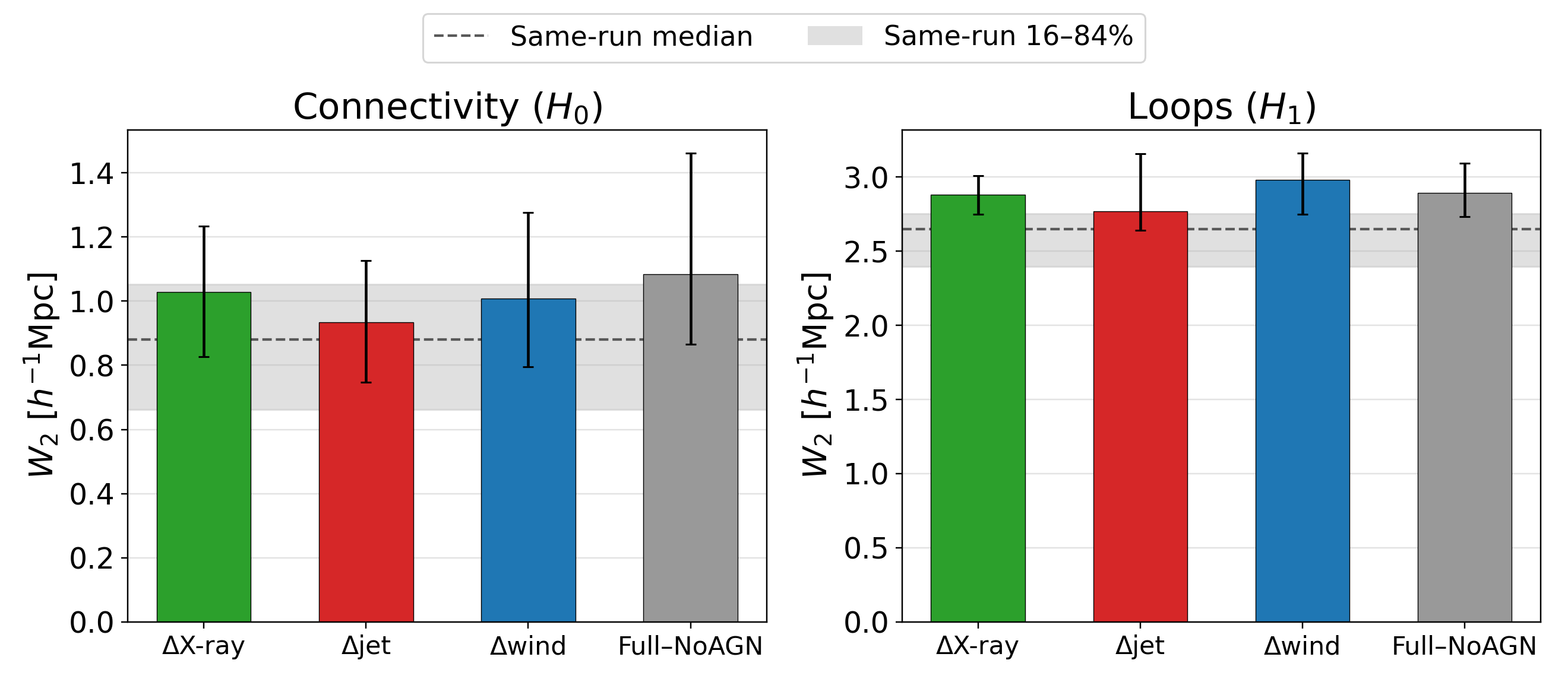}
\caption{\label{fig:total_transport} Median 2-Wasserstein distance for the X-ray, jet, and wind feedback steps (Eq.~\ref{eq:diff_homology}), plus a fourth bar for the direct Full Physics-to-No-AGN comparison, separately for $H_0$ and $H_1$, for the total galaxy sample. Bars report median distances, error bars span the 16th--84th percentiles of the 50 subsampling iterations, and the gray band marks the Full Physics same-run reference. The total galaxy sample shows little feedback-dependent difference in $H_0$ connectivity and only a marginal excess above the same-run reference in the characteristic scales of $H_1$ features. All three adjacent $H_1$ steps exceed the Full Physics same-run reference, albeit only marginally for the jet step, as does the direct Full Physics-to-No-AGN comparison in both $H_0$ and $H_1$, although Appendix~\ref{app:reference_sensitivity} shows that this result is more reference-dependent than the satellite-galaxy comparison.}
\end{figure}

Figure~\ref{fig:total_transport} shows the median 2-Wasserstein distance for the X-ray, jet, and wind feedback steps defined in equation~\ref{eq:diff_homology}, together with a fourth bar giving the direct Full Physics-to-No-AGN comparison, separately for $H_0$ and $H_1$, for the total galaxy sample. Bars report median distances, error bars span the 16th--84th percentiles of the 50 subsampling iterations, and the gray band marks the Full Physics same-run reference. For the total galaxy sample, the nearest-neighbor connectivity changes little among feedback realizations, while the characteristic scales of loop-like galaxy configurations show a modest and reference-dependent difference.
The three adjacent-step $H_0$ distances remain largely within the same-run sampling reference, consistent with similar local connectivity. In $H_1$, all three adjacent steps rise slightly above the Full Physics reference band, though by a much smaller margin for the jet step than for X-ray or wind. The direct Full Physics-to-No-AGN comparison exceeds the same-run reference in both $H_0$ and $H_1$, consistent with the cumulative effect of the three adjacent steps; for $H_0$ in particular, this cumulative exceedance is not visible in any of the three individual steps on their own. Because this classification depends more strongly on the reference realization than the satellite result below (see Appendix~\ref{app:reference_sensitivity}), we treat the full galaxy sample differences as suggestive rather than significant. The absence of a clear enhancement in the full-sample Betti curves further indicates that the models do not simply differ in the total number of large-scale loop-like configurations. Instead, their persistence diagrams are suggestive of slight shifts in the birth and death scales of finite-persistence loop features.

We also apply the same test to halos. To make this comparison, the halo-center catalog replaces every halo by a single point which is the minimum of the potential, thereby removing the internal satellite configuration while retaining the nodal and filamentary distribution of host halos \cite{bond1996filaments,cooray2002halo}. This is nearly equivalent to considering only central galaxies and ignoring all satellites (modulo that central galaxies do not live exactly at, albeit usually close to, the halo minimum potential). Appendix~\ref{app:central_transport} repeats this transport diagnostic directly on the central-galaxy catalog and finds a similarly small feedback-dependent signal, consistent with this near-equivalence.

\begin{figure}[!htbp]
\centering
\includegraphics[width=\columnwidth]{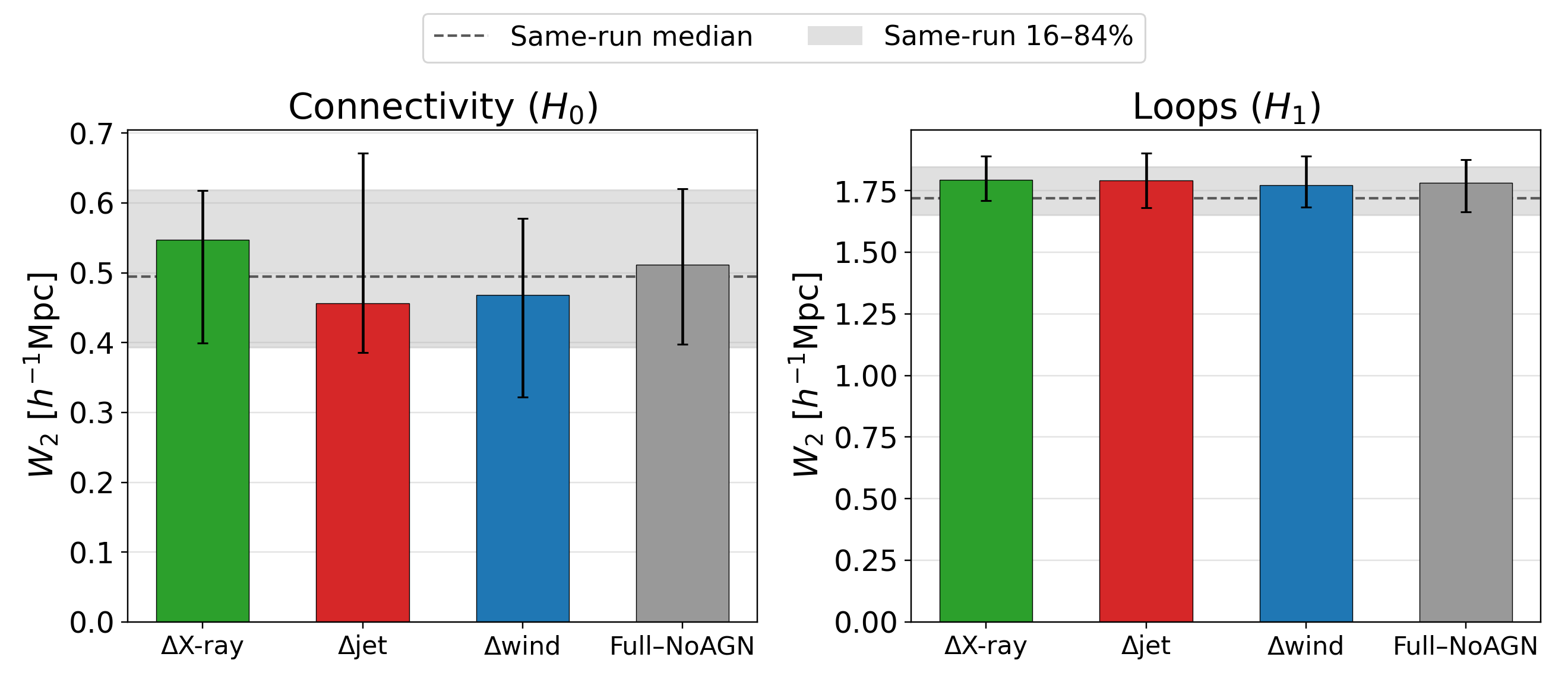}
\caption{\label{fig:halo_center_transport} Median 2-Wasserstein distance for the X-ray, jet, and wind feedback steps, plus a fourth bar for the direct Full Physics-to-No-AGN comparison, separately for $H_0$ and $H_1$, for the halo-center catalog, following the same convention as Fig.~\ref{fig:total_transport}: bars report median distances, error bars span the 16th--84th percentiles of the 50 subsampling iterations, and the gray band marks the Full Physics same-run reference. The large-scale distribution of halo centers shows no comparable difference above the same-run sampling reference, including for the direct Full Physics-to-No-AGN comparison.}
\end{figure}

Figure~\ref{fig:halo_center_transport} shows the same median 2-Wasserstein distance diagnostic as Fig.~\ref{fig:total_transport}, including the direct Full Physics-to-No-AGN bar, but for the halo-center catalog. The halo-center $W_2$ distances, including the direct Full Physics-to-No-AGN comparison, remain within the Full Physics reference band. This test shows that the feedback-dependent differences are more closely associated with the locations of galaxies within and around halos rather than with a large displacement of the halo centers themselves, as expected since the dark matter halo locations should not be very sensitive to baryonic feedback effects.  Appendix~\ref{app:reference_sensitivity} confirms that this result is independent of which {\sc Simba} variant is used as the reference.

\section{FEEDBACK DIFFERENCES IN SATELLITE GALAXIES}\label{sec:results2}

\subsection{Satellite population 2-Wasserstein distances}

The small but non-trivial loop differences for the full galaxy sample versus none for the halo sample hints that topological differences among AGN feedback variants may be more pronounced among the satellite galaxy populations.  Here we will demonstrate that this is indeed the case, and explore how satellite topology varies with AGN feedback.

\begin{figure}[t]
\centering
\includegraphics[width=\columnwidth]{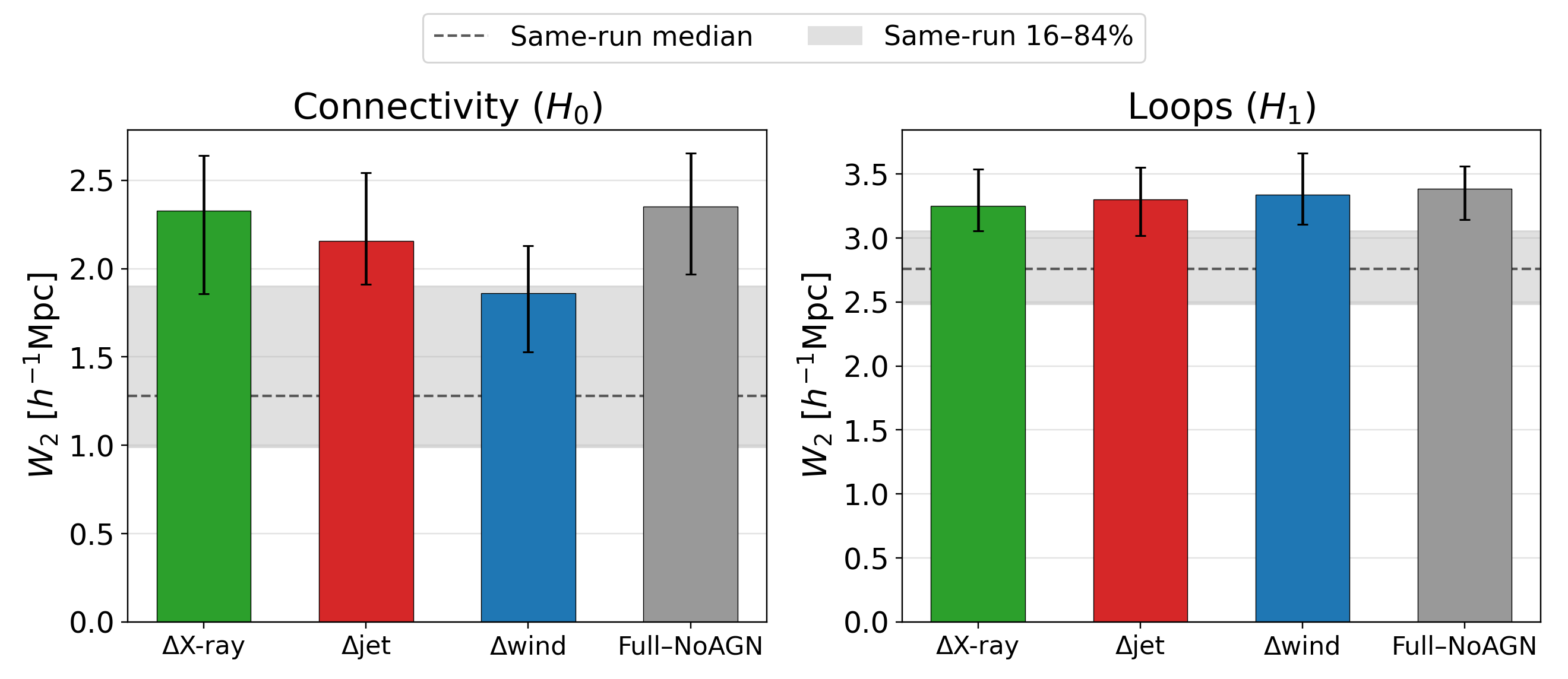}
\caption{\label{fig:satellite_transport} Median 2-Wasserstein distance for the X-ray, jet, and wind feedback steps, plus a fourth bar for the direct Full Physics-to-No-AGN comparison, separately for $H_0$ and $H_1$, for the satellite galaxy sub-population. Bars show median distances, error bars show the 16th--84th percentiles, and the gray band shows the Full Physics same-run reference. Satellite galaxies show the clearest feedback-dependent differences in $H_1$: the median 2-Wasserstein distance exceeds the Full Physics same-run reference for all three adjacent feedback steps, and the direct Full Physics-to-No-AGN comparison exceeds it by the widest margin of the four bars in both $H_0$ and $H_1$.}
\end{figure}

In Fig.~\ref{fig:satellite_transport} we show the median 2-Wasserstein distance for the X-ray, jet, and wind feedback steps, together with a fourth bar for the direct Full Physics-to-No-AGN comparison, separately for $H_0$ and $H_1$, considering only the satellite galaxy sub-population. Bars show median distances, error bars show the 16th--84th percentiles, and the gray band shows the Full Physics same-run reference.
Both the connectivity and loop persistence diagrams show clear differences among the models. For $H_0$, there is a large difference between the Full-Physics run and the No-Xray run, a sizeable difference again when turning off jets, while turning off radiative AGN winds does not exceed the same-run reference.  Meanwhile for $H_1$, in each case turning off a feedback module results in a difference that exceeds the same-run reference in the satellite topology.  The direct Full Physics-to-No-AGN comparison exceeds the same-run reference by the widest margin of all four bars in both $H_0$ and $H_1$, consistent with the cumulative effect of the three adjacent steps.  This result is stable to the choice of reference run (Appendix~\ref{app:reference_sensitivity}): for all choices, all differences between feedback variants continue to exceed the same-run reference.

It may be counter-intuitive that satellites are more susceptible to AGN feedback, when the primary purpose of AGN feedback in simulations is to quench massive central galaxies.  But in terms of topology, satellites are more strongly affected by environmental processes, including filamentary preprocessing \citep{bulichi2024galaxy}, so their spatial distribution can be altered more easily by AGN feedback.  More interesting is that all forms of AGN feedback seem to have some influence, it is not restricted to only AGN jets even though jets are primarily responsible for quenching \citep{scharre2024effects}.


\subsection{The Scale Dependence of Satellite Differences}

The $W_2$ distance and its diagonal/off-diagonal decomposition established above that the satellite $H_0$/$H_1$ differences exceed the same-run reference and that they mainly reflect shifts in the birth and death scales of topological features, rather than the gain or loss of persistent features through point-to-diagonal matching. Because $W_2$ summarizes each pairwise comparison between two persistence diagrams as a single number, however, it does not by itself indicate at which filtration scales $\alpha$ those shifts occur. To localize these differences to specific scale ranges, we return to the Betti curves, which are computed explicitly as a function of $\alpha$ and can therefore be inspected directly for where the feedback realizations diverge.

Satellite galaxies show fairly consistent differences in the topology of their loop configurations across feedback realizations.  Here we investigate the dependence of these differences on filtration scale $\alpha$. This is to contrast with the two-point correlation function of the total galaxy sample (Sec.~\ref{sec:results1}), which shows differences confined to sub-Mpc scales; Appendix~\ref{app:satellite_2pcf} repeats this same-tracer comparison directly on the satellite catalog used here.

\begin{figure*}[t]
\centering
\includegraphics[width=0.9\textwidth]{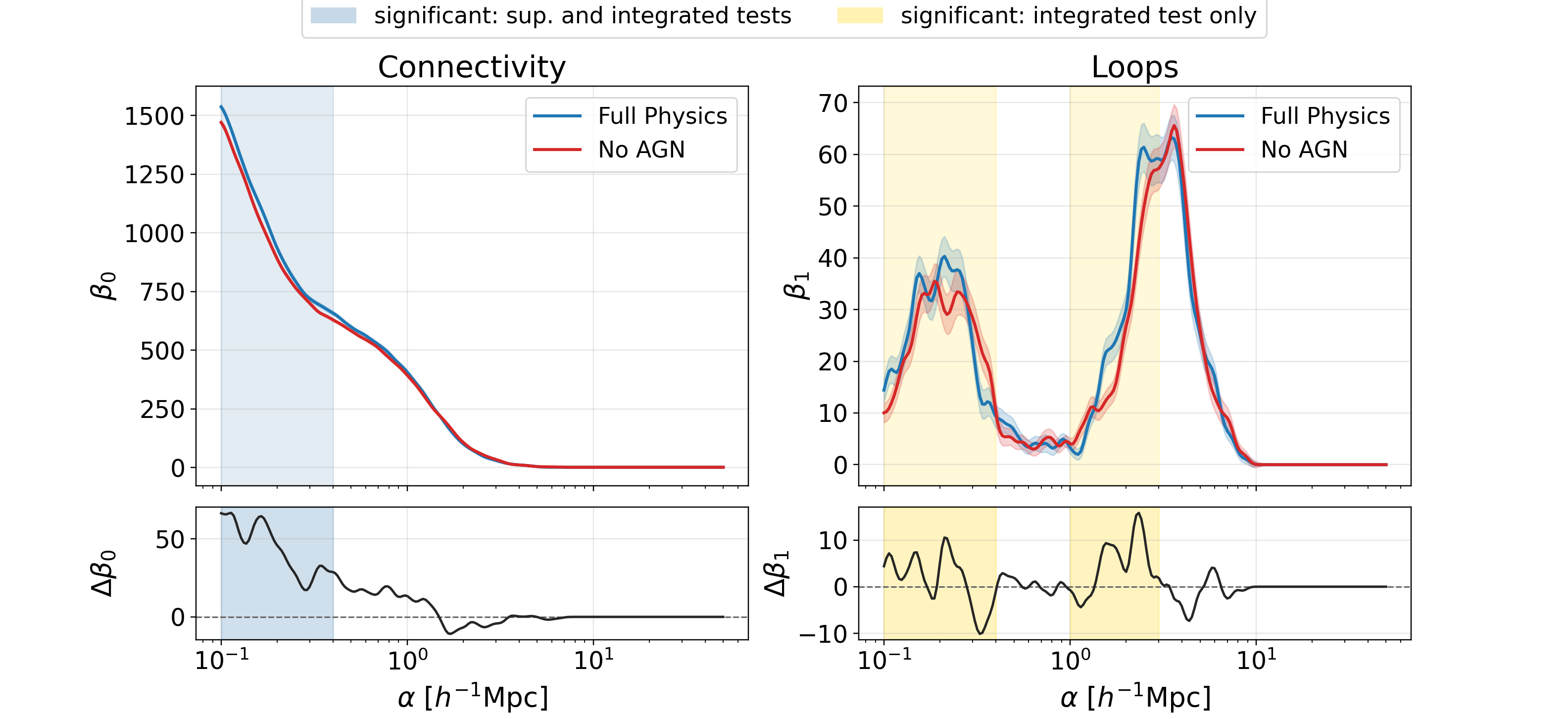}
\caption{\label{fig:satellite_betti} Betti numbers $\beta_0$ (left) and $\beta_1$ (right) vs. filtration scale $\alpha$ for satellite galaxies, comparing Full Physics and No-AGN; each panel has an attached sub-panel below showing the difference $\Delta\beta_k(\alpha)$. Shaded vertical bands mark the two tested scale windows, $\alpha\sim0.1$--$0.4\,h^{-1}\mathrm{Mpc}$ (halo-internal/circum-halo) and $\alpha\sim1$--$3\,h^{-1}\mathrm{Mpc}$ (outer-halo/inter-halo), color-coded by which functional test detects a significant difference there for this comparison (Sec.~\ref{sec:methodology} and Sec.~\ref{sec:results2}): blue where both the supremum ($D_{\max}$) and integrated ($T_{\rm area}$) statistics are significant, gold where only the integrated statistic is significant, and unshaded where neither is. The $\beta_0$ response is confined to the lower window; the $\beta_1$ response is significant in both windows, but only in the integrated statistic, indicating a broad, coherent separation rather than a localized deviation.}
\end{figure*}

Figure~\ref{fig:satellite_betti} shows the satellite Betti numbers $\beta_0$ and $\beta_1$ as a function of filtration scale $\alpha$, comparing Full Physics and No-AGN directly; each panel has an attached sub-panel below showing the difference $\Delta\beta_k(\alpha)$. Two scale ranges show a visible separation between the curves: $\alpha\sim0.1$--$0.4\,h^{-1}\mathrm{Mpc}$, associated mainly with halo-internal and circum-halo satellite configurations, and $\alpha\sim1$--$3\,h^{-1}\mathrm{Mpc}$, associated with outer-halo and inter-halo configurations.

We quantify these differences using the $D_{\max}$ and $T_{\rm area}$ statistics defined in Sec.~\ref{sec:methodology}. For each statistic and homology dimension, we test four model comparisons (the three adjacent feedback steps and the direct Full Physics-to-No-AGN comparison) over three scale ranges: the full filtration range and the two windows above. This gives $4\times3=12$ tests per statistic and homology dimension; because testing many comparisons increases the chance of a spuriously significant result, we apply Holm--Bonferroni correction \cite{holm1979simple} across each set of 12.

For loops ($H_1$), the lower-scale window shows a robust response: $T_{\rm area}$ is significant for all four comparisons ($p_{\rm Holm}=0.002$--$0.026$), and $D_{\max}$ is significant for all three adjacent feedback steps (X-ray, jet, and wind; $p_{\rm Holm}=0.002$--$0.038$), though not for the direct Full Physics-to-No-AGN comparison ($p_{\rm Holm}=0.29$). At $\alpha\sim1$--$3\,h^{-1}\mathrm{Mpc}$, none of the individual feedback steps is significant, whereas the direct Full Physics-to-No-AGN comparison is significant in $T_{\rm area}$ ($p_{\rm Holm}=0.009$) but not in $D_{\max}$ ($p_{\rm Holm}=0.056$), indicating a broad cumulative response rather than a localized deviation.

For connectivity ($H_0$), significant differences are confined to the $0.1$--$0.4\,h^{-1}\mathrm{Mpc}$ window; no comparison is significant at $\alpha\sim1$--$3\,h^{-1}\mathrm{Mpc}$. All other tested comparisons are non-significant after the Holm correction.

Further investigation reveals that the satellite differences mainly reflect shifts in the characteristic scales and persistence of loop-like galaxy configurations, rather than a large change in their total number. This interpretation comes from noting that across all three feedback steps, $96$--$99\%$ of the $W_2$ cost is carried by off-diagonal point-to-point matches rather than point-to-diagonal matches, indicating that the diagram-level difference is dominated by shifts in the birth and death scales of loop-like features, rather than by the gain or loss of persistent loop features through point-to-diagonal matching. 

Overall, we have identified satellite galaxies as a population in which the feedback-dependent spatial differences are clearer than when (also) considering centrals. But satellites also tend to be lower mass and more quenched, so this effect may be a manifestation of stellar mass or star formation rate differences.  In the next section we investigate topological differences as a function of these quantities. 

\section{DEPENDENCE ON HALO MASS AND STAR-FORMATION RATE}\label{sec:results3}

Having found that satellite galaxies show the clearest spatial differences, we now ask where related differences appear in the total galaxy population. We compare galaxies in fixed host-halo-mass bins and then compare star-forming and quenched galaxies, always at fixed galaxy number density.

\subsection{Halo Mass Dependence}
\label{sec:activation_scales}

\begin{figure*}[t]
\centering
\includegraphics[width=0.85\textwidth]{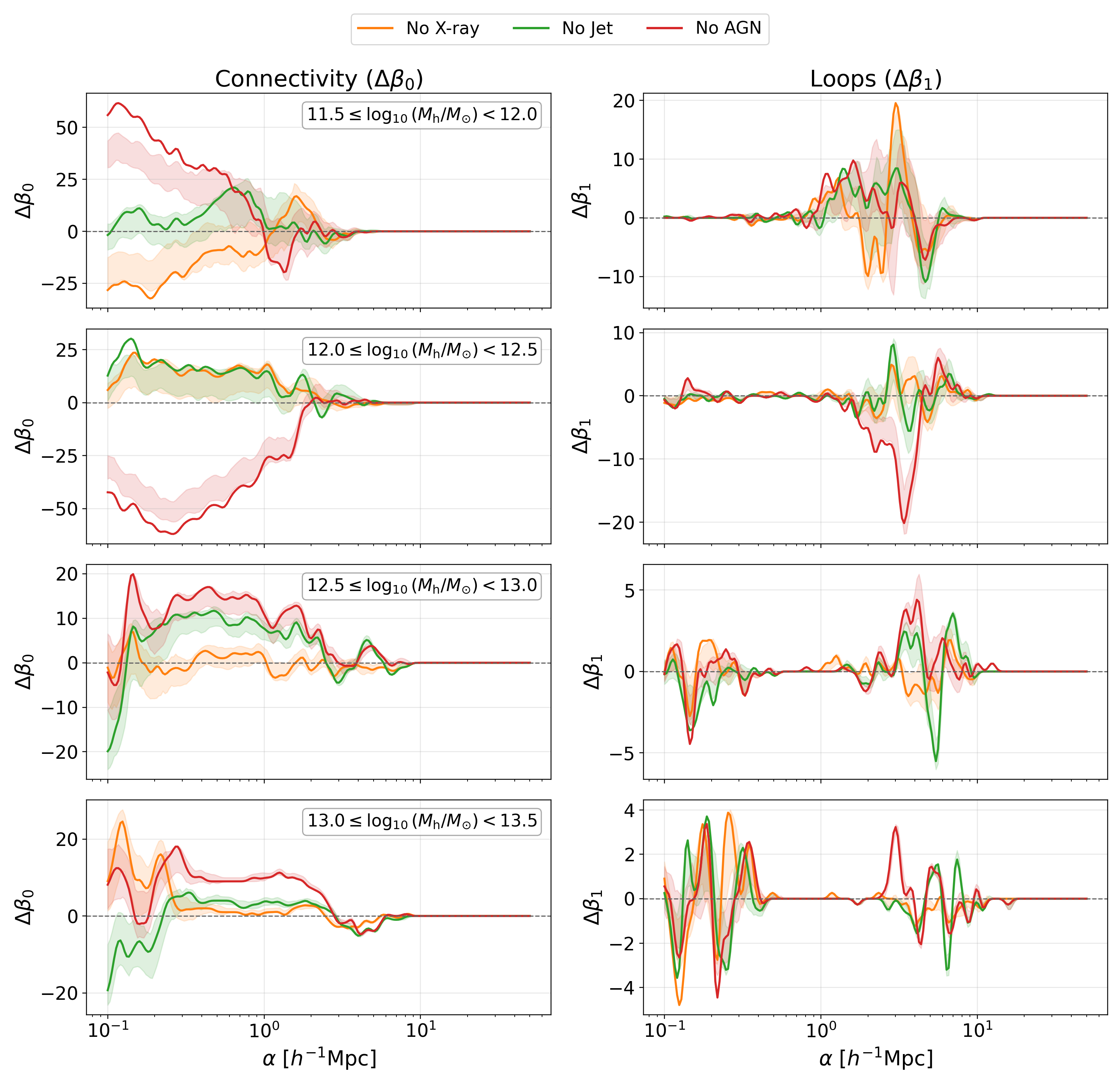}
\caption{\label{fig:exp4_betti_fixed_mh} Betti-number differences relative to Full Physics, $\Delta\beta_0=\beta_0(\mathrm{run})-\beta_0(\mathrm{Full\ Physics})$ and $\Delta\beta_1=\beta_1(\mathrm{run})-\beta_1(\mathrm{Full\ Physics})$ (left/right columns), vs. filtration scale $\alpha$, for the No-Xray, No-Jet, and No-AGN {\sc Simba} runs (see legend), in the four host-halo-mass bins from Table~\ref{tab:sample_summary}, labeled inside each left-column panel (rows, top to bottom: bins 1-4). The dashed line marks $\Delta\beta=0$ (i.e. equal to Full Physics), and the shaded bands show the 16th--84th percentile envelope of $\Delta\beta$ from the same repeated $80\%$ subsampling used throughout the paper. Host-mass bin 0 ($11.0\leq\log_{10}(M_{\mathrm{h}}/M_{\odot})<11.5$) shows no visible feedback-dependent separation in the Betti curves and is therefore omitted, though Appendix~\ref{app:reference_sensitivity} finds only a marginal, reference-dependent $W_2$ exceedance in this bin. Feedback-dependent $H_1$ differences become visible near $M_{\mathrm{h}}\sim10^{12}M_{\odot}$ and extend across two characteristic scale ranges at higher host masses. The onset of the separation coincides with the regime where efficient AGN kinetic feedback becomes important in {\sc Simba}.}
\end{figure*}

As in Sec.~\ref{sec:results2}, we use Betti curves rather than the aggregate $W_2$ distance here because they retain the $\alpha$-axis needed to identify the halo-mass regimes and scales at which feedback differences emerge. Fig.~\ref{fig:exp4_betti_fixed_mh} shows the Betti-number differences relative to Full Physics, $\Delta\beta_0$ and $\Delta\beta_1$, as a function of filtration scale $\alpha$ for the No-Xray, No-Jet, and No-AGN {\sc Simba} runs, subdivided into four halo mass bins summarized in Table~\ref{tab:sample_summary}.  For both Betti numbers, the two lowest-mass bins shown, $11.5\leq\log_{10}(M_{\mathrm{h}}/M_{\odot})<12.5$, show comparably large spread among the curves. For connectivity, the No-AGN run lies well below the other curves for $\alpha<1$~Mpc (i.e. $\Delta\beta_0$ is strongly negative) in the $12.0\leq\log_{10}(M_{\mathrm{h}}/M_{\odot})<12.5$ bin, and the difference is primarily driven by radiative AGN feedback which, when included, strongly increases $\beta_0$ in this regime.  For loops, the differences in this same bin appear at intermediate scales of $\alpha\sim 2-4$~Mpc, where again radiative AGN feedback causes a pronounced jump in $\Delta\beta_1$.  This radiative-AGN-driven pattern weakens in the next higher mass bin. In the highest mass bin, the dominant feedback channel depends on which homology dimension is considered: for loops ($H_1$), the dominant driver shifts to the jet feedback step, consistent with the corresponding $W_2$ decomposition in Appendix~\ref{app:fixed_mass_bin_wasserstein}, whereas for connectivity ($H_0$), the radiative-AGN (wind) step remains the largest contributor.

It is particularly interesting that radiative AGN feedback seems to most strongly impact topology in moderately massive halos.   {\sc Simba}'s radiative feedback model has minimal impact on galaxy quenching, which is primarily driven instead by AGN jet feedback at least for central galaxies.  Radiative AGN feedback occurs at high Eddington ratios (above $\sim 0.1$), which tend to occur in gas-rich galaxies that are still forming stars.  It appears that turning on this feedback mode has a strong impact in changing the spatial distribution of satellites within $\sim 10^{12} M_\odot$ halos.  This may be because in more massive halos, satellite galaxies can be stripped of gas by other means such as ram pressure stripping, but in moderate mass halos the satellites can still have ongoing star formation that can be altered by radiative AGN feedback.

\subsection{Star-Forming and Quenched Samples}

\begin{figure*}[t]
\centering
\includegraphics[width=0.95\textwidth]{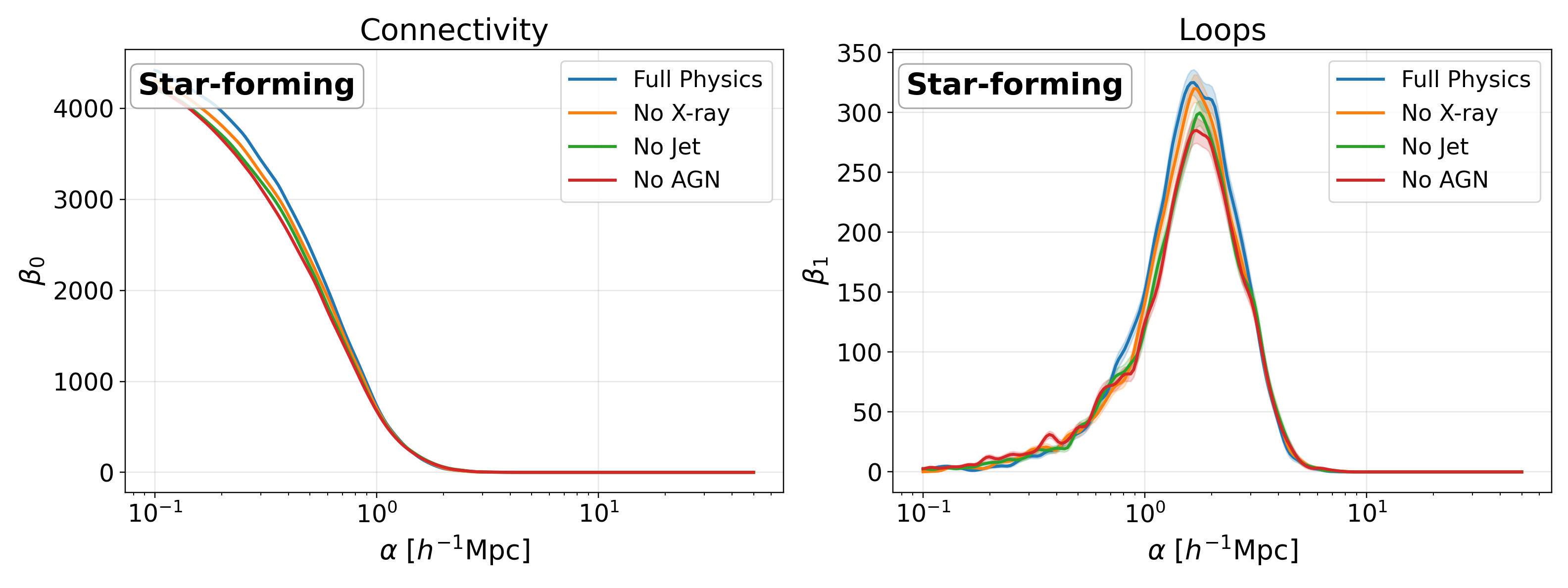}\\[8pt]
\includegraphics[width=0.95\textwidth]{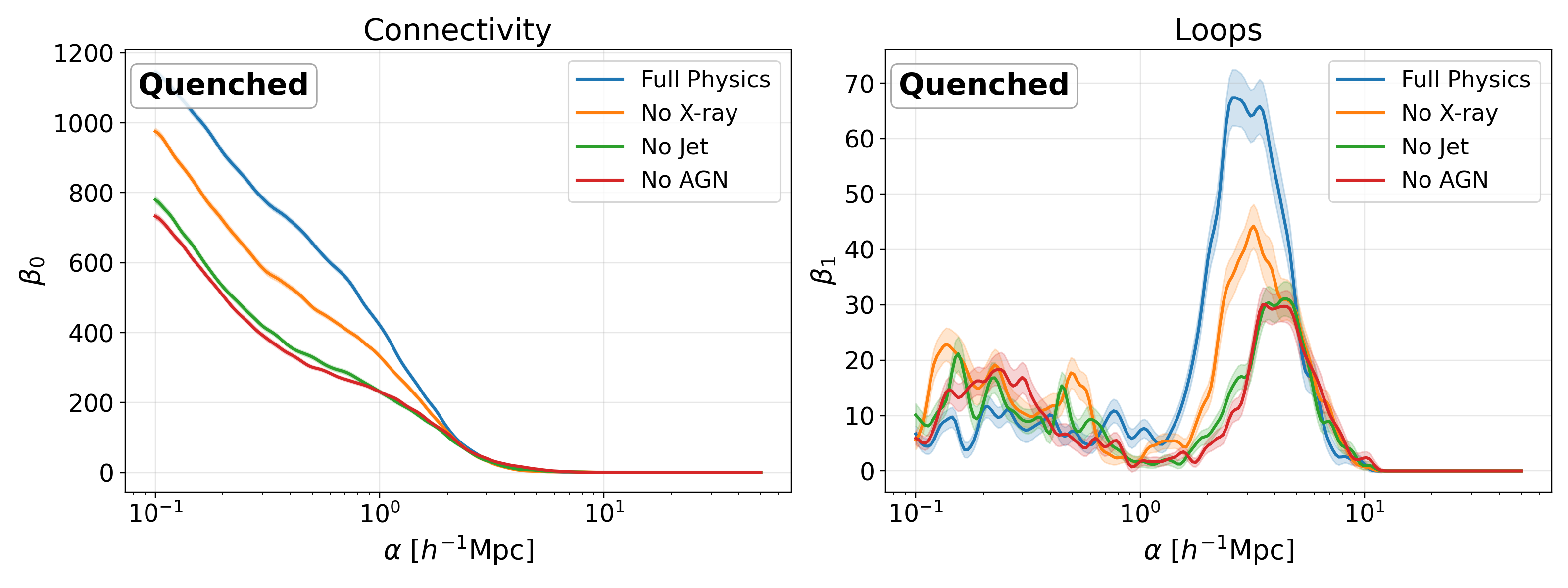}
\caption{\label{fig:quenched_absolute} Betti curves $\beta_0(\alpha)$ and $\beta_1(\alpha)$ for the star-forming (top panels) and quenched (bottom panels) samples, at fixed galaxy number density for each of the four feedback realizations. Quenched galaxies show clearer feedback-dependent differences than star-forming galaxies: the quenched sample has an approximately ordered small-scale $H_0$ difference and a higher Full Physics $H_1$ curve on multi-virial to inter-halo scales, while the star-forming curves show weaker and less structured separation. These differences are not caused by the total number of selected galaxies, but they can include both changes in central--satellite composition and differences in the spatial distribution of the quenched population.}
\end{figure*}

For the same reason as in Sec.~\ref{sec:results2}, we again inspect Betti curves here rather than the aggregate $W_2$ distance, since they show directly on which scales the quenched and star-forming samples differ. Figure~\ref{fig:quenched_absolute} compares $\beta_0(\alpha)$ and $\beta_1(\alpha)$ for the quenched and star-forming samples at fixed galaxy number density, with the star-forming sample shown in the top panels and the quenched sample in the bottom panels.

The star-forming Betti curves remain relatively close across feedback realizations in both connectivity and loop-like features, with no strongly ordered separation, whereas the quenched galaxies show a structured model dependence. At small filtration scales, the quenched $H_0$ curves are approximately ordered as
$\mathrm{Full\ Physics}>\mathrm{No\mbox{-}Xray}>\mathrm{No\mbox{-}Jet}>\mathrm{No\mbox{-}AGN}$.
For the same number of rank-selected quenched galaxies, this means that the Full Physics realization contains more disconnected quenched components at small $\alpha$, while the No-AGN realization contains fewer components at the same scale.

The quenched galaxies also differ on LSS filament scales. At $\alpha\sim1.5$--$5\,h^{-1}\mathrm{Mpc}$, the Full Physics $H_1$ curve lies above the weaker-feedback realizations, corresponding to a larger number of live loop-like configurations in this scale range. This visible Betti-curve separation occurs on the same broad scales identified in the satellite and fixed-host-mass comparisons. It remains a count-level signature of model-dependent galaxy configurations, not a direct measurement of individual loop radii or representative cycles.


This interpretation requires one important caveat. Fixed galaxy number density does not mean fixed central--satellite composition. As shown in Appendix~\ref{app:composition}, the satellite fraction of the quenched sample decreases from $f_{\rm sat}^{\rm Q}=0.894$ in No-AGN to $0.504$ in Full Physics, whereas the star-forming satellite fraction remains much more stable, $f_{\rm sat}^{\rm SF}=0.238$--$0.258$. Part of the quenched-galaxy difference therefore reflects the changing mixture of central and satellite galaxies. Central galaxies sit close to their host halo's density peak, essentially one per halo, whereas satellites of the same halo are typically found close to one another; a satellite-dominated sample therefore merges into connected components at smaller $\alpha$ than a central-dominated one. A more central-rich quenched sample can affect the small-scale $H_0$ ordering and alter which galaxies within and between halos contribute to $H_1$ features.

The central--satellite mixture does not exhaust the comparison. Among quenched satellites alone, the median halo-centric radius is $R/R_{200c}=0.674$ in No-AGN and $1.061$ in Full Physics (Appendix~\ref{app:composition}). This is a genuine positional difference independent of the central--satellite mixture discussed above: even restricted to satellites alone, quenched satellites in Full Physics typically sit near or beyond the virial radius, whereas those in No-AGN typically sit well within it. Satellites farther from the halo center are more plausibly positioned to participate in the outer-halo and inter-halo loop-like configurations discussed in Sec.~\ref{sec:results2}, so the quenched $H_1$ difference coincides with a change in where quenched satellites reside within and around their host halos, although the present analysis does not isolate the quantitative contribution of the radial difference to the topology.

Taken together, the sample comparisons show that feedback-dependent spatial differences are clearest for satellites, galaxies in massive halos, and quenched galaxies. The quenched comparison does not isolate a pure spatial effect at fixed central--satellite composition, but it identifies both the changing satellite fraction and the halo-centric distribution as relevant accompanying differences.

\section{DISCUSSION}\label{sec:discussion}

The sample comparisons consistently show that the clearest feedback-dependent spatial differences occur among satellite galaxies, galaxies in massive halos, and quenched galaxies. The large-scale distribution of halo centers shows no comparable difference, and central galaxies remain closer to the same-run sampling reference than satellites. The satellite differences occur on halo-internal/circum-halo scales, and, in the cumulative feedback contrast, on outer-halo/inter-halo scales as well. This combination of population, environment, and scale dependence of topology constitutes ``spatial quenching bias.'' 

\subsection{Physical Scenario for Spatial Quenching Bias}

The quenched populations differ in both central--satellite composition and halo-centric distribution. At fixed galaxy number density, the quenched sample is more central-rich in Full Physics than in No-AGN, which can affect the small-scale connectivity ordering and change which galaxies contribute to loop configurations. The difference is not limited to this changing mixture: among quenched satellites alone, the median radius is $R/R_{200c}=0.674$ in No-AGN and $1.061$ in Full Physics, while the host-halo-mass distributions remain broadly similar. The feedback realizations therefore differ in the locations of quenched satellites within and around their halos as well as in the satellite fraction of the quenched sample. These accompanying differences provide the concrete basis for the spatial-quenching-bias interpretation.

Within {\sc Simba}, these spatial differences occur in the halo-mass regime where kinetic AGN feedback is expected to be important for quenching massive galaxies, although the measured galaxy distribution reflects the combined action of multiple feedback steps. {\sc Simba} studies find that quenched systems inhabit hot, baryon-depleted circumgalactic gas, consistent with cumulative starvation after energetic bipolar feedback \cite{dave2019simba,appleby2021low,appleby2020impact,yang2024feedback}. Satellite quenching may also be delayed relative to the initial loss of environmental gas supply, allowing satellites to continue forming stars over an orbital or depletion interval, while backsplash populations can return toward halo outskirts after pericentric passage \cite{oman2016satellite,wetzel2014galaxy,gill2005evolution}. One possible physical picture is therefore that delayed-quenching or backsplash populations contribute galaxies near and beyond $R_{200c}$, increasing the number of outer-halo and inter-halo configurations traced at high $\alpha$. This picture is consistent with the measured radial difference and the enhanced quenched $H_1$ curve, but it remains a hypothesis rather than a causal explanation demonstrated by the TDA measurements.


\subsection{Physical Scales of the Satellite Differences}
\label{sec:physical_calibration}

The two scale ranges identified for satellites correspond approximately to different halo environments. Halos in the relevant mass range, from $M_{200c}\sim10^{12}M_{\odot}$ to representative group-scale systems with $\log_{10}(M_{200c}/M_{\odot})\approx13.1$, have characteristic radii of order $R_{200c}\sim0.15$--$0.4\,h^{-1}\mathrm{Mpc}$. The lower range, $\alpha\sim0.1$--$0.4\,h^{-1}\mathrm{Mpc}$, therefore probes halo-internal and circum-halo galaxy configurations, including sub-virial, near-virial, and few-virial-radius scales depending on host mass; as shown in Sec.~\ref{sec:results2}, this response is robust, detected in the integrated statistic for all four feedback comparisons and in the supremum statistic for all three adjacent steps. The upper range, $\alpha\sim1$--$3\,h^{-1}\mathrm{Mpc}$, extends well beyond individual halo radii and is sensitive to outer-halo, inter-halo, and bridge-like configurations; here, however, only the cumulative Full Physics-to-No-AGN comparison reaches significance, and only in the integrated statistic, indicating a broad, coherent shift rather than a localized deviation, and one that is not individually resolved in any single feedback step. This comparison is an order-of-magnitude calibration rather than a one-to-one conversion from filtration scale to physical loop radius: an $H_1$ feature is supported by multiple galaxies, edges, and higher-dimensional simplices, so its birth and death scales do not directly measure the radius of a single structure.

These two scale ranges are not equally visible in all halos. The feedback-dependent $H_1$ separation begins around $M_{200c}\sim10^{12}M_{\odot}$ and becomes clearer at higher host masses (Sec.~\ref{sec:activation_scales}), qualitatively aligning it with the regime where efficient AGN kinetic feedback and massive-halo quenching become important in {\sc Simba}. The result therefore describes a spatial pattern associated with massive host halos rather than a generic feature of the full galaxy population.

This scale interpretation remains phenomenological. The Betti curves identify where in scale and host mass the galaxy distributions differ, and the Wasserstein distances show that these differences exceed the same-run sampling variation. They do not determine whether the underlying cause is gas heating, a change in satellite orbits and survival, halo-centric redistribution, altered galaxy formation histories, or a combination of these processes.

\subsection{What PD/$W_2$ Adds Beyond Betti Curves and Pair Statistics}

The methodological value of the PD/$W_2$ analysis is not that it detects a signal entirely absent from the 2PCF. Rather, the global 2PCF identifies an overall scale-dependent clustering response in the untrimmed galaxy catalogs, whereas the controlled PD/$W_2$ comparisons identify the galaxy populations, host environments, homology dimensions, and filtration scales in which feedback-dependent spatial differences are clearest. A Betti curve records only how many generators are alive at each filtration scale, whereas a persistence diagram retains the birth--death coordinates of those generators. Consequently, two catalogs can have similar live $H_1$ counts while still differing in the characteristic birth scales, death scales, and persistence of their loop-forming configurations.

This distinction is essential for interpreting the total-galaxy and satellite results together. The models can contain similar numbers of live $H_1$ features while differing in the spatial scales and persistence of the galaxy configurations that support them. This difference is suggestive in the total-galaxy sample and more robust among satellites. In persistence-diagram terms, the satellite $W_2$ cost is dominated by off-diagonal matching, indicating that the birth and death scales of loop-like features shift more than they are gained or lost through point-to-diagonal matching. Physically, this suggests that AGN feedback primarily reorganizes the characteristic spatial scales of loop-forming galaxy configurations, rather than simply increasing or decreasing the number of persistent loop-like structures. This does not track individual real-space loops, but it identifies population-level and topological-scale structure that is not resolved by the global two-point function or absolute Betti curves.

The controlled sample comparisons show where this additional information arises. Halo centers test the large-scale distribution of host halos, the central--satellite comparison identifies satellites as the clearest population, fixed host-mass bins place the onset near $10^{12}M_{\odot}$, and the star-forming/quenched comparison shows a more structured difference among quenched galaxies. PD/$W_2$ therefore complements rather than replaces Betti curves and two-point statistics by measuring scale-dependent differences that those summaries compress.

\subsection{Scope and Limitations}

Several limitations should be kept explicit. First, this study uses a single matched-phase {\sc Simba} volume at $z=0$. The matched initial conditions are ideal for isolating feedback-induced differences, but they do not by themselves measure the full cosmic-variance distribution of the statistic. Our same-run reference-realization test checks whether the finite-sampling reference band depends on which feedback realization is used to construct it; it does not estimate the variance among independent cosmological volumes. Testing the full finite-volume and cosmic-variance sensitivity of the persistence-diagram statistic therefore remains a task for larger-volume or multi-seed simulation suites. In addition, although the underlying {\sc Simba} simulation is periodic, our alpha complexes are constructed in one Euclidean fundamental domain rather than on a three-dimensional torus. Connections across opposite box faces are therefore omitted. The identical box geometry and matched phases make part of this boundary contribution common to the four realizations, and the primary $H_0$ and $H_1$ interpretation concerns scales much smaller than the box size, but neither consideration guarantees exact cancellation in the inter-model persistence-diagram distances. A genuinely periodic alpha-complex calculation is an important future robustness test. Second, the fixed-number-density construction controls the leading sampling-density degeneracy, but it does not control every population property, including the central--satellite composition of a rank-selected sample, and it is not equivalent to a full observational selection function. Redshift-space distortions, survey masks, fiber collisions or blending, stellar-mass uncertainties, and photometric-redshift errors may all affect persistence diagrams and must be included in future mock-observation tests.

Third, the interpretation of individual generators has limited uniqueness. Persistent homology gives a stable global summary of connected components, loops, and voids, but a single generator can admit multiple representative cycles. For this reason, the statistical claim rests on ensemble PD/$W_2$ distances, same-run finite-sampling references, fixed-host-mass total-galaxy comparisons, and controlled population splits rather than on the identification of individual generators with unique real-space counterparts. Consequently, our static $z=0$ analysis measures the final geometric configuration of the tracer populations, but does not trace the dynamical formation history of these topological features. Tracking the physical expansion of specific loops, the orbital histories or phase-space evolution of individual satellites, or the formation history of the quenched population would require full merger-tree analyses in future work.

As a separate methodological point, Appendix~\ref{app:inverse_mapping} gives a proof-of-concept visualization of inverse mapping from a local $H_1$ generator to a real-space galaxy configuration. That example is included only to illustrate a possible future diagnostic and is not used as evidence for the ensemble-level physical interpretation in this paper.

\subsection{Implications for Future Observations}

For observational cosmology, a practical advantage of PD/$W_2$ is that it identifies the galaxy populations, halo masses, topological dimensions, and filtration-scale ranges in which spatial differences are clearest. In the present analysis, the 2PCF already detects a small-scale feedback response, while PD/$W_2$ further localizes the response to satellite and quenched-galaxy distributions within and around massive halos.

This observational direction is further motivated by the reduced-density satellite check in Appendix~\ref{app:top50_masscut}. When the sample is restricted to the most massive 50\% of satellites, raising the effective minimum stellar mass by nearly an order of magnitude, the satellite $H_1$ difference remains above the broadened same-run sampling reference and the two characteristic scale ranges remain visible. The result is therefore not driven solely by the faintest satellites and may remain informative under stricter stellar-mass completeness limits. This is still only a simulation-level approximation to a mass-limited observational sample; realistic survey masks, redshift uncertainties, blending, and group-finder errors must be included in mock-observation tests.

If this behavior persists in larger volumes and realistic mock observations, the method should be useful for comparisons among simulation suites and ultimately for survey constraints on baryonic feedback. A practical next step is to reproduce the comparisons used here---all galaxies, halo or group centers, central and satellite galaxies, fixed host-mass bins, and star-forming and quenched populations---in mock surveys. Such tests should include survey masks and window functions, redshift-space distortions or photometric-redshift smearing, stellar-mass completeness and ranking scatter, fiber collisions or blending, and central--satellite misclassification by group finders. They should also determine whether the persistence-diagram distances remain complementary to two-point statistics after these observational and finite-volume effects are included.

\section{CONCLUSION}\label{sec:summary}

We summarize our main results in the order they appear (Secs.~\ref{sec:results1}--\ref{sec:results3}):

\begin{itemize}
\item \textbf{Two-point clustering (Fig.~\ref{fig:2pcf}):} feedback-dependent differences in the galaxy two-point correlation function are largest on sub-Mpc scales and converge to relative differences of order $10\%$ or below beyond a few $h^{-1}\mathrm{Mpc}$.
\item \textbf{Global Betti curves (Fig.~\ref{fig:global_topology}):} the absolute Betti curves for the total galaxy sample change only weakly among the matched {\sc Simba}-50 feedback realizations, showing no large volume-averaged topological difference and motivating the transport-based comparisons that follow.
\item \textbf{Total-galaxy and halo-center transport (Figs.~\ref{fig:total_transport}, \ref{fig:halo_center_transport}):} the total galaxy sample shows only a marginal, reference-dependent excess in loop-like transport cost, whereas the halo-center catalog shows no comparable difference above the same-run sampling reference, and central galaxies remain closer to that reference than satellites.
\item \textbf{Satellite transport (Fig.~\ref{fig:satellite_transport}):} satellite galaxies show the clearest feedback-dependent spatial differences, with all three adjacent feedback comparisons, and the direct Full Physics-to-No-AGN comparison by the widest margin, exceeding the same-run reference in loop-like features.
\item \textbf{Scale dependence of the satellite response (Fig.~\ref{fig:satellite_betti}):} the Betti curve differences are robust at intra-halo scales ($\alpha\sim0.1$--$0.4\,h^{-1}\mathrm{Mpc}$) for every individual feedback step, while at LSS scales ($\alpha\sim1$--$3\,h^{-1}\mathrm{Mpc}$) it is statistically evident only in the cumulative Full Physics-to-No-AGN comparison as a broad, coherent shift rather than a sharply localized one. Optimal matching shows that off-diagonal point-to-point matches account for $96.5$--$99.0\%$ of this satellite Wasserstein cost, indicating that the diagram-level difference is dominated by scale shifts in the birth and death scales of loop-like features, rather than by the gain or loss of persistent loop features through point-to-diagonal matching.
\item \textbf{Halo-mass dependence (Fig.~\ref{fig:exp4_betti_fixed_mh}):} related differences become visible in the total galaxy population near host masses of $M_{\mathrm{h}}\sim10^{12}M_{\odot}$, with radiative AGN feedback the dominant driver in this moderately massive-halo regime.
\item \textbf{Star-forming vs.\ quenched galaxies (Fig.~\ref{fig:quenched_absolute}):} the topological differences are more clearly structured among quenched than star-forming galaxies, coinciding with both a changing central--satellite mixture and a larger median $R/R_{200c}$ for quenched satellites in Full Physics than in No-AGN.
\end{itemize}

The TDA signal is therefore most closely associated with satellite galaxies, massive halos, and the locations of quenched galaxies within and between halos. Although the TDA measurements do not establish a unique causal mechanism, we suggest that AGN feedback from massive galaxies alters the satellite distribution by moving baryons to larger scales, which shifts the birth and death scales of loop topological configurations. Persistent homology and optimal transport thus complement the small-scale feedback dependence already visible in the 2PCF by highlighting its population, environment, and topological-scale structure.

\begin{acknowledgments}
RRK acknowledges support from the THERS Make New Standards Program for the Next Generation Researchers. The authors thank Shohei Kanda for developing preliminary analysis scripts during the early stages of this project. 
TTT has been supported by JSPS Grant-in-Aid in Scientific Research (24H00247), JST CREST Grant No.\ JPMJCR24Q1, and by the Joint Research program of the Institute of Statistical Mathematics (General Research 2) entitled ``Machine-Learning Cosmogony: From Structure Formation to Galaxy Evolution.''
HS is supported by the National SKA Program of China (No. 2020SKA0110401), NSFC (Grant No. 12103044), Yunnan Provincial Key Laboratory of Survey Science with project (No. 202449CE340002) and Yunnan Revitalization Talents Support Plan.
The authors used OpenAI ChatGPT (GPT-5.6-Sol), Anthropic Claude (Claude Sonnet 5, including Claude Code), and Google Gemini (Gemini 3.1 Pro) as AI-assisted tools during manuscript preparation, including language editing, scientific discussion of the results, and verification of citations, cross-references, and reported figures against the underlying analysis outputs. These tools were directed interactively through natural-language instructions, and all AI-generated or AI-suggested text, numerical claims, and citations were independently checked by the authors against the analysis code and cached results before inclusion. See Sec.~\ref{sec:methodology} for the disclosed use of AI-assisted coding tools in the analysis pipeline itself. The authors reviewed and take full responsibility for all scientific analyses, interpretations, and conclusions in this paper.
\end{acknowledgments}

\section*{Data Availability}
The {\sc Simba} simulation data, the analysis code and derived persistence-diagram data underlying this study are available from the corresponding author upon reasonable request.

\clearpage

\appendix

\section{Implementation Notes}
Persistence diagrams are computed from $\alpha$-complex filtrations using the GUDHI library \cite{gudhi:urm}. Although the underlying {\sc Simba} simulation uses periodic boundary conditions, we construct the standard Euclidean GUDHI alpha complex from the galaxy coordinates in a single fundamental domain. We do not use a periodic Delaunay triangulation, replicate the point set across box faces, or impose minimum-image connections in the filtration. Consequently, galaxies close across opposite faces of the simulation box are not joined through the boundary. This choice avoids the duplicate simplices and persistence generators that would arise from naively analyzing a replicated point set, but it can introduce boundary effects. Because the four feedback realizations have matched phases, identical box geometry, and the same coordinate boundaries, part of this contribution is common to their paired contrasts; moreover, our primary $H_0$ and $H_1$ interpretation concerns scales of a few $h^{-1}\mathrm{Mpc}$, well below the $50\,h^{-1}\mathrm{Mpc}$ box size. These facts do not ensure exact cancellation, however, and we treat a genuinely periodic alpha-complex calculation as a future robustness test.

GUDHI stores $\alpha$-complex filtration values as squared radii. Before constructing Betti curves, Wasserstein inputs, and all figures in this paper, we convert the finite birth and death values to the corresponding linear radius scale by taking their square roots; the plotted and quoted $\alpha$ values are therefore in $h^{-1}\mathrm{Mpc}$, not in squared-distance units. The $W_2$ distances are evaluated using \texttt{gudhi.wasserstein.wasserstein\_distance} \cite{gudhi:WassersteinDistance}, with \texttt{order=2} and \texttt{internal\_p=2} throughout. Although this GUDHI routine supports essential points with infinite coordinates, we replace the infinite death coordinate of the single essential $H_0$ class with a common finite placeholder of $1000.0$ before transport evaluation. The essential-generator count is fixed across the diagrams entering each like-for-like comparison: $H_0$ has one essential class, corresponding to the final connected component, and $H_1$ and $H_2$ have none for the completed finite $\alpha$-complex. We check this count before transport evaluation. Because all like-for-like diagrams contain exactly one essential $H_0$ class with the same birth coordinate, this replacement contributes zero transport cost; the reported distances are instead controlled by finite-scale generators. For the subsampling analysis, random seeds are fixed to ensure reproducibility across the $N_{\mathrm{iter}}=50$ repeated subsampling realizations. The physical interpretation in the main text is drawn from these ensemble transport distributions together with the corresponding Betti-curve shifts in controlled subpopulations.

\section{Reference-Realization Sensitivity of the Same-Run Finite-Sampling Reference}\label{app:reference_sensitivity}
The main transport figures use the same-run finite-sampling reference band constructed from the Full Physics realization. To test whether the inferred exceedance depends on this reference choice, we repeated the same construction using each feedback realization as the reference catalog. For each sample, homology dimension, and adjacent feedback step, we asked whether the median inter-model $W_2$ distance exceeds the 84th percentile of the same-run reference distribution constructed from Full Physics, No-Xray, No-Jet, or No-AGN.

\begin{figure*}[t]
\centering
\includegraphics[width=0.8\textwidth]{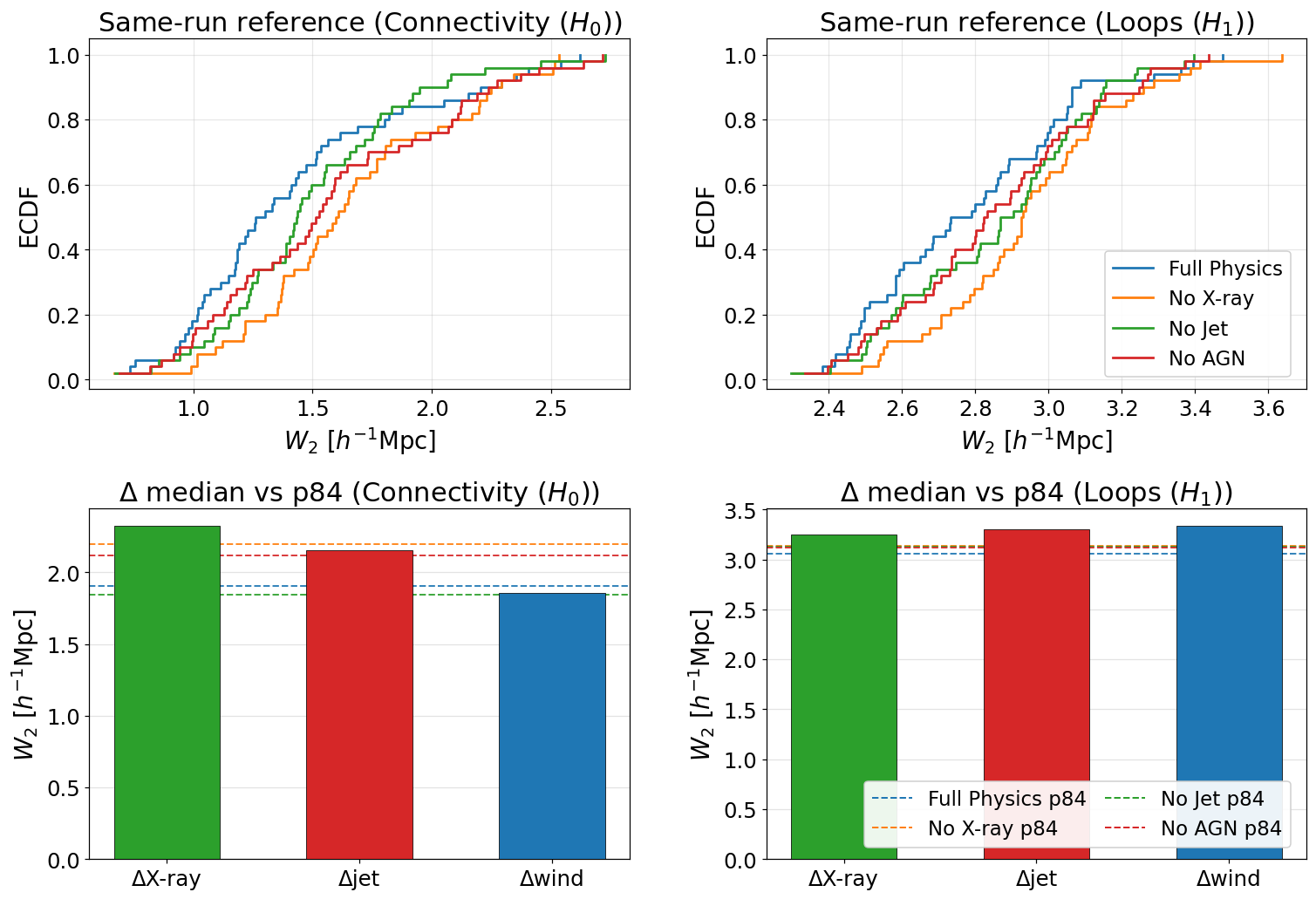}
\caption{\label{fig:reference_sensitivity_satellite} Reference-realization sensitivity of the same-run finite-sampling reference for the satellite catalog. Upper panels show the empirical cumulative distributions of same-run $W_2$ distances constructed from each feedback realization; lower panels compare adjacent feedback-step medians with the corresponding 84th percentiles. For $H_1$, all three feedback-step medians exceed every reference choice, with modest margins of $3.6$--$9.4\%$; this is a finite-sampling robustness criterion rather than a formal significance level.}
\end{figure*}

Figure~\ref{fig:reference_sensitivity_satellite} shows the reference-realization sensitivity for the satellite catalog. The upper panels show the empirical cumulative distributions of the same-run reference distances for the four possible reference realizations. The lower panels compare the adjacent feedback-step medians with the corresponding 84th-percentile reference values. In $H_1$, all three feedback-step medians exceed the 84th-percentile reference for every reference realization. This confirms that the primary satellite-loop result is not driven by adopting the Full Physics run as the reference catalog. In $H_0$, the classification is more channel-dependent, so we treat the satellite $H_0$ response as secondary.

Table~\ref{tab:reference_sensitivity_summary} summarizes the same test for the main diagnostics and gives the satellite-$H_1$ fractional margins. Panel A shows that the total-galaxy $H_1$ exceedance depends on which feedback realization is used to construct the same-run reference: the median exceeds the reference for some choices but not for others. In contrast, the halo-center $H_1$ control remains below the corresponding reference threshold for every reference choice and every adjacent feedback step. The fixed-host-mass bins show that the most robust $H_1$ response occurs in bins 2--4, consistent with the mass scale identified in the Betti-curve analysis. The quenched and star-forming population splits both show several exceedances, so the population-level interpretation is based on the shape and scale dependence of the Betti curves, not on a binary exceedance criterion alone. Panel B shows that all satellite-$H_1$ margins are positive but modest, ranging from $3.6\%$ to $9.4\%$.

\begin{center}
\begingroup
\scriptsize
\setlength{\tabcolsep}{2.1pt}
\renewcommand{\arraystretch}{0.86}
\refstepcounter{table}\label{tab:reference_sensitivity_summary}
\begin{minipage}{0.98\columnwidth}
\raggedright
\textbf{TABLE~\thetable.} Reference-realization sensitivity of the same-run finite-sampling reference. Panel A lists median-exceedance classifications for the main diagnostics. Each four-character entry uses the reference order (Full Physics, No-Xray, No-Jet, No-AGN), with Y/N indicating whether the median adjacent-feedback $W_2$ distance exceeds the corresponding same-run 84th percentile. Panel B gives satellite-$H_1$ fractional margins, $m = (\widetilde W_2^\Delta - Q_{0.84}(W_2^{\rm same}))/Q_{0.84}(W_2^{\rm same})$.
\end{minipage}

\vspace{0.2em}

\textit{Panel A: Median-exceedance classification.}

\vspace{0.1em}

\resizebox{\columnwidth}{!}{%
\begin{tabular}{llccc}
\hline
Sample & Homology & $\Delta_{\rm X-ray}$ & $\Delta_{\rm jet}$ & $\Delta_{\rm wind}$ \\
\hline
Total galaxies        & $H_1$ & YNYY & YNYY & YNYY \\
Halo centers          & $H_1$ & NNNN & NNNN & NNNN \\
Centrals              & $H_1$ & YNYY & YNYY & NNNY \\
Satellites            & $H_0$ & YYYY & YNYY & NNYN \\
Satellites            & $H_1$ & YYYY & YYYY & YYYY \\
\hline
Host-mass bin 0       & $H_1$ & YNYY & YNYY & YNYY \\
Host-mass bin 1       & $H_1$ & YNYY & YNYN & YNYY \\
Host-mass bin 2       & $H_1$ & YYYY & YYYY & YYYY \\
Host-mass bin 3       & $H_1$ & YYYY & YYYY & YYYY \\
Host-mass bin 4       & $H_1$ & YYYY & YYYY & YYYY \\
\hline
Quenched galaxies     & $H_1$ & YYYY & YYYY & YYYY \\
Star-forming galaxies & $H_1$ & YYYY & YYNY & YYNY \\
\hline
\end{tabular}%
}

\vspace{0.2em}
\vspace{\baselineskip}

\textit{Panel B: Fractional margins for the satellite $H_1$ diagnostic.}

\vspace{0.1em}

\resizebox{\columnwidth}{!}{%
\begin{tabular}{lcccc}
\hline
Channel & Full Physics ref. & No-Xray ref. & No-Jet ref. & No-AGN ref. \\
\hline
$\Delta_{\rm X-ray}$ & $+6.5\%$ & $+3.6\%$ & $+3.8\%$ & $+4.1\%$ \\
$\Delta_{\rm jet}$   & $+8.1\%$ & $+5.2\%$ & $+5.4\%$ & $+5.7\%$ \\
$\Delta_{\rm wind}$  & $+9.4\%$ & $+6.4\%$ & $+6.7\%$ & $+7.0\%$ \\
\hline
\end{tabular}%
}

\vspace{0.15em}

\begin{minipage}{\columnwidth}
\raggedright
Note. This table summarizes median exceedance only; it does not require the full 16--84\% inter-model interval to lie above the reference band. The complete set of $H_0$ and $H_1$ exceedance classifications is available from the corresponding author upon reasonable request.
\end{minipage}
\endgroup
\end{center}

\newpage
\section{Central-Galaxy Tracer Control}\label{app:central_transport}
Figure~\ref{fig:central_transport_appendix} shows the same 2-Wasserstein diagnostic for the central-galaxy catalog. This control complements the satellite-only result in Fig.~\ref{fig:satellite_transport}. The central-galaxy transport costs remain close to the same-run finite-sampling reference, especially compared with the systematic satellite $H_1$ exceedance. This supports the interpretation that the satellite-$H_1$ catalog is the clearer tracer-type diagnostic, without requiring the stronger statement that satellites uniquely carry all of the feedback response.

\begin{center}
\includegraphics[width=0.96\columnwidth]{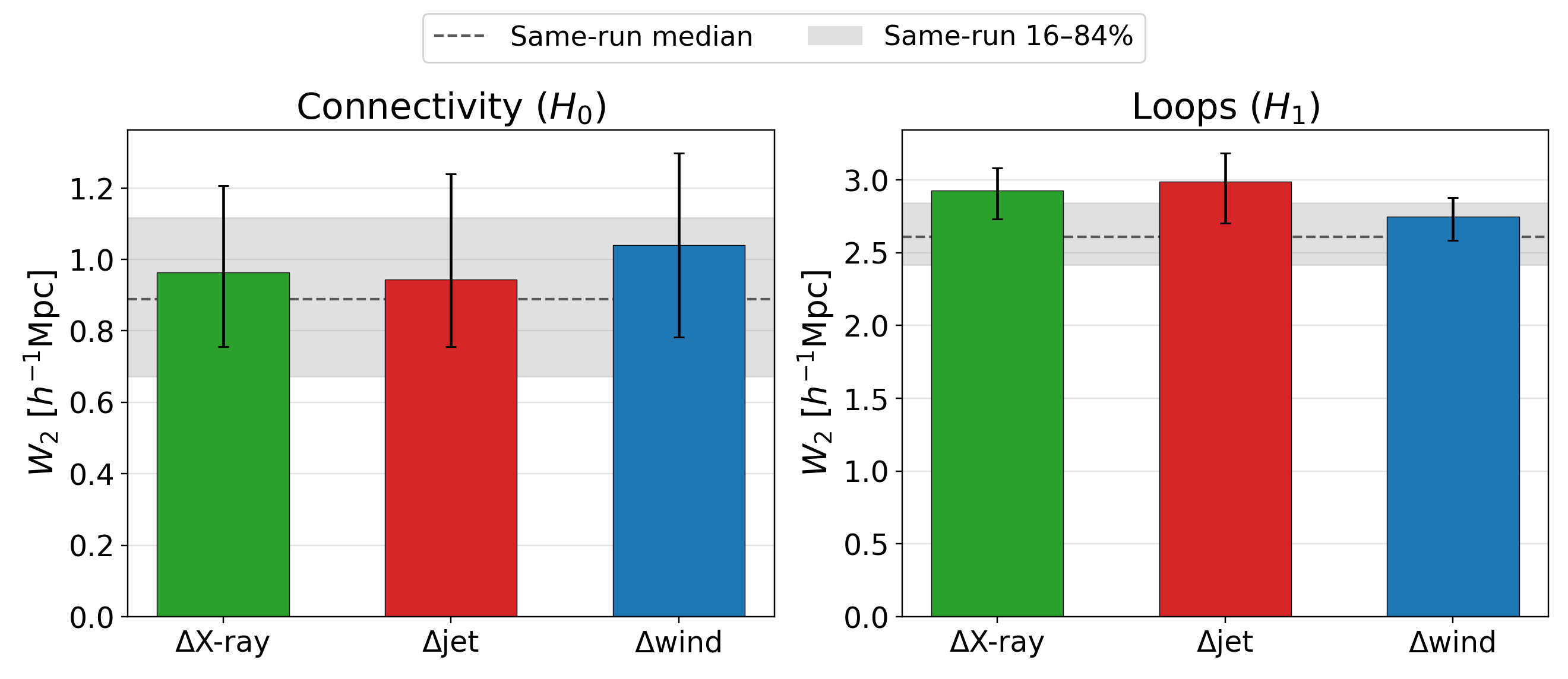}
\refstepcounter{figure}\label{fig:central_transport_appendix}
\vspace{0.1em}
\begin{minipage}{0.96\columnwidth}
\footnotesize
\textbf{FIG.~\thefigure.} 2-Wasserstein distances for the central-galaxy catalog. The bars report the median transport cost, the error bars span the 16th to 84th percentiles, and the gray shaded band marks the same-run finite-sampling reference. The central-galaxy response remains close to the reference relative to the satellite-$H_1$ diagnostic in Fig.~\ref{fig:satellite_transport}.
\end{minipage}
\end{center}

\section{Fixed Host-Halo-Mass-Bin Wasserstein Distances}\label{app:fixed_mass_bin_wasserstein}
The main text shows the fixed-host-mass Betti curves for the total-galaxy catalog because they provide the clearest visual representation of the host-mass dependence of the two-branch response. Figure~\ref{fig:fixed_mass_bin_wasserstein_appendix} gives the corresponding $W_2$ decomposition for the same displayed bins, with a fourth bar in each panel for the direct Full Physics-to-No-AGN comparison, following the same convention as Figs.~\ref{fig:total_transport}--\ref{fig:satellite_transport}. These bar plots are the quantitative transport-cost counterpart to Fig.~\ref{fig:exp4_betti_fixed_mh}: the difference persists after controlling for host halo mass, and in the $H_1$ sector the dominant feedback channel changes with mass, with the wind step more prominent in the lower-mass bins and the jet step overtaking the wind step in bin4. The direct Full Physics-to-No-AGN comparison exceeds the same-run reference in every displayed bin and in both $H_0$ and $H_1$, consistent with the cumulative effect of the three adjacent steps. In bin1's connectivity channel, the direct comparison is the only one of the four bars to exceed the reference at all, while none of the three individual steps does. The single largest exceedance margin of any bar in any bin is the direct comparison in bin4's connectivity channel, where the median exceeds the reference by more than a factor of four; for loops in bin4, however, the jet step exceeds the reference by a wider margin than the direct comparison does.

\begin{widetext}
\vspace{0.2em}
\centering
\begin{minipage}{0.47\textwidth}
\centering
{\scriptsize\textbf{bin 1:} $11.5\leq\log_{10}(M_{\mathrm{h}}/M_{\odot})<12.0$}\\[5pt]
\includegraphics[width=\linewidth]{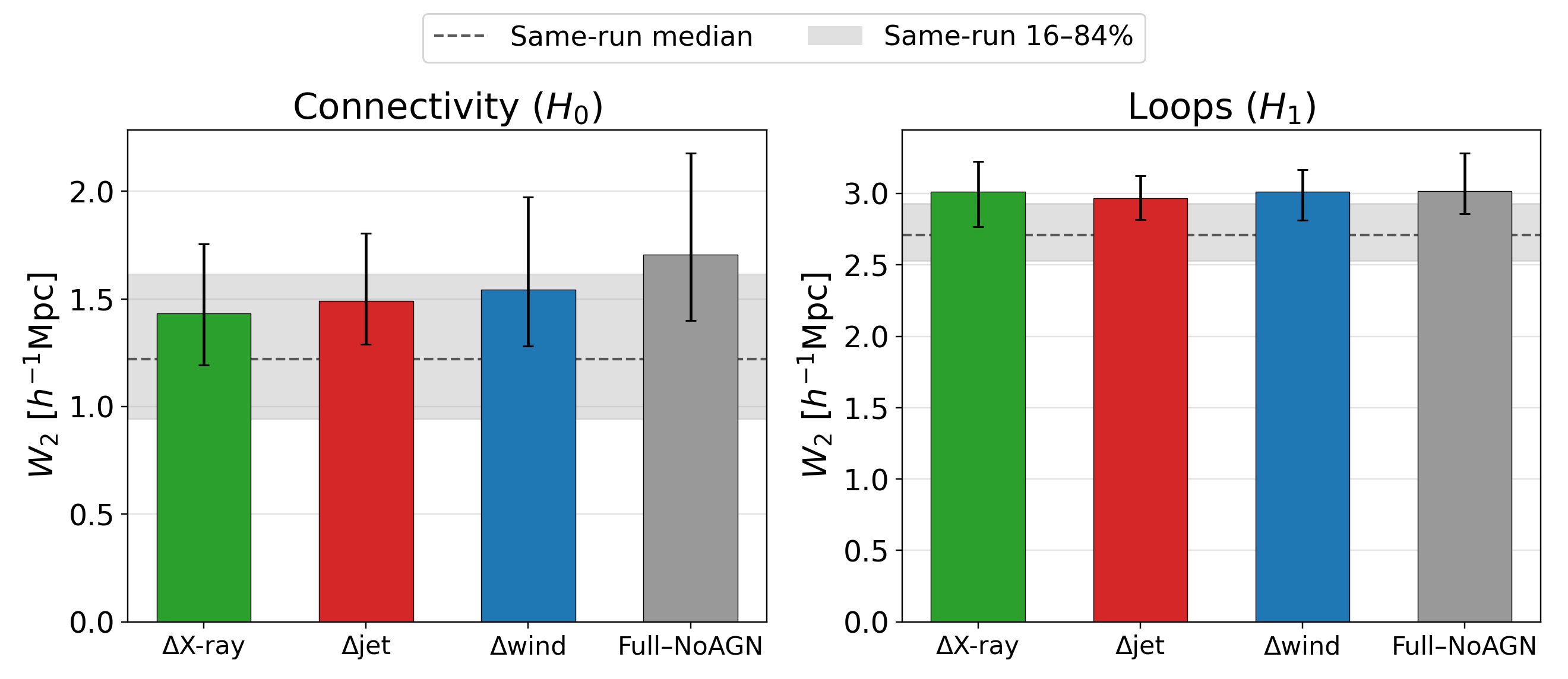}
\end{minipage}\hfill
\begin{minipage}{0.47\textwidth}
\centering
{\scriptsize\textbf{bin 2:} $12.0\leq\log_{10}(M_{\mathrm{h}}/M_{\odot})<12.5$}\\[5pt]
\includegraphics[width=\linewidth]{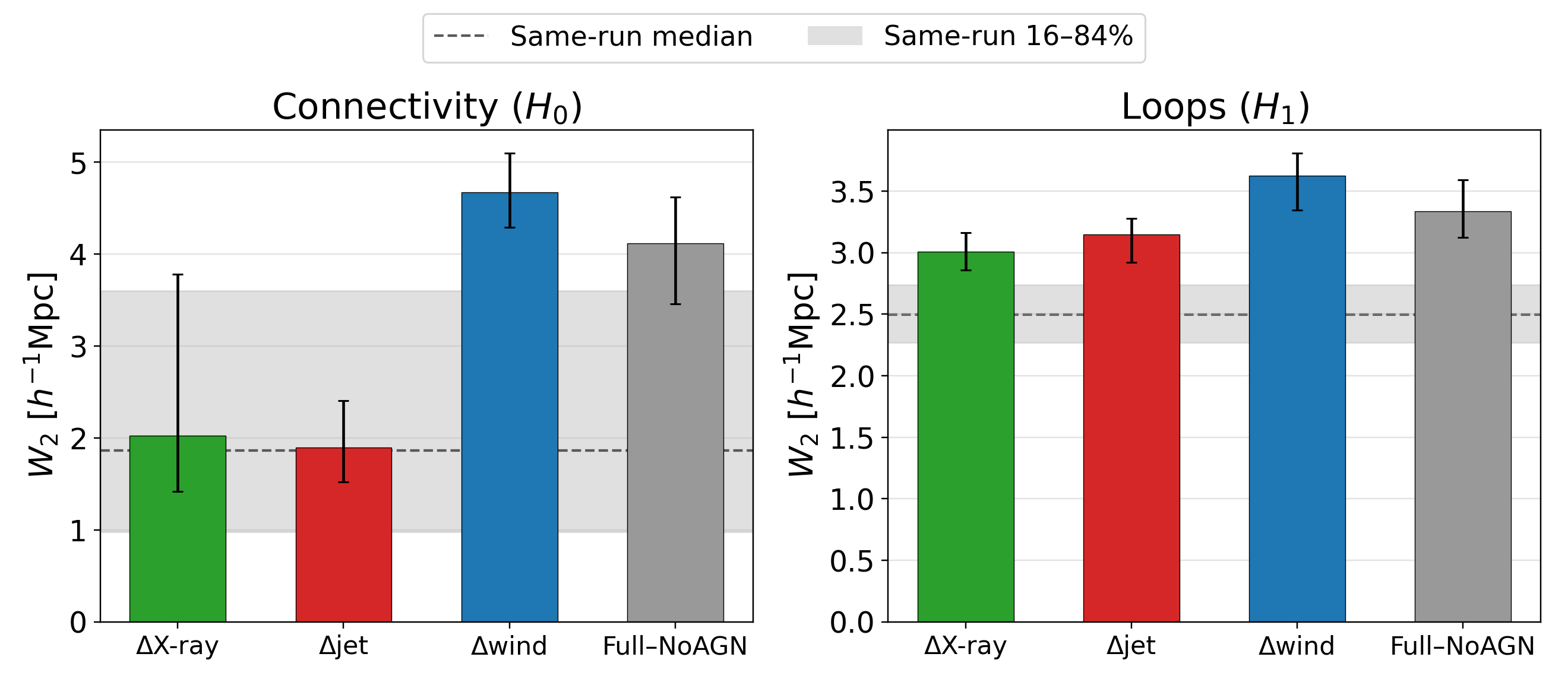}
\end{minipage}\\[3pt]
\begin{minipage}{0.47\textwidth}
\centering
{\scriptsize\textbf{bin 3:} $12.5\leq\log_{10}(M_{\mathrm{h}}/M_{\odot})<13.0$}\\[5pt]
\includegraphics[width=\linewidth]{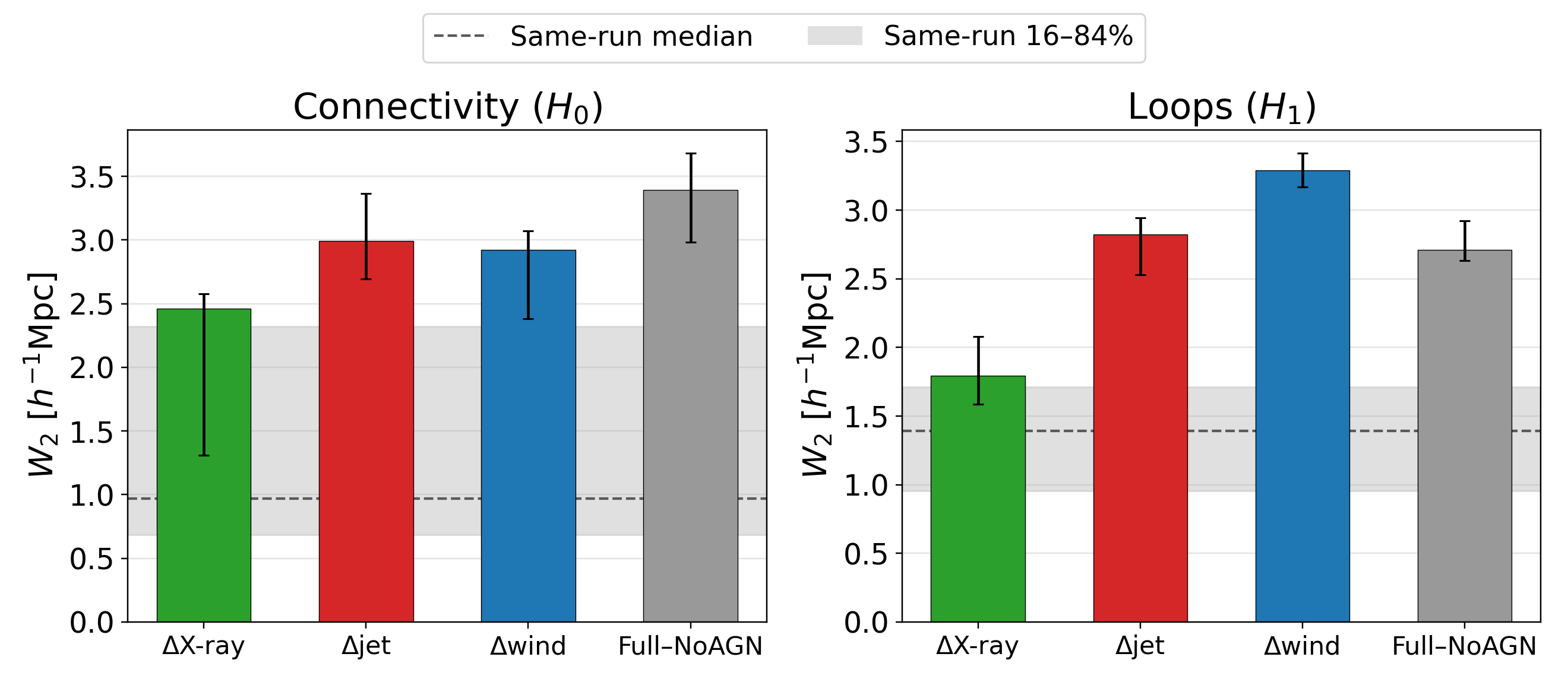}
\end{minipage}\hfill
\begin{minipage}{0.47\textwidth}
\centering
{\scriptsize\textbf{bin 4:} $13.0\leq\log_{10}(M_{\mathrm{h}}/M_{\odot})<13.5$}\\[5pt]
\includegraphics[width=\linewidth]{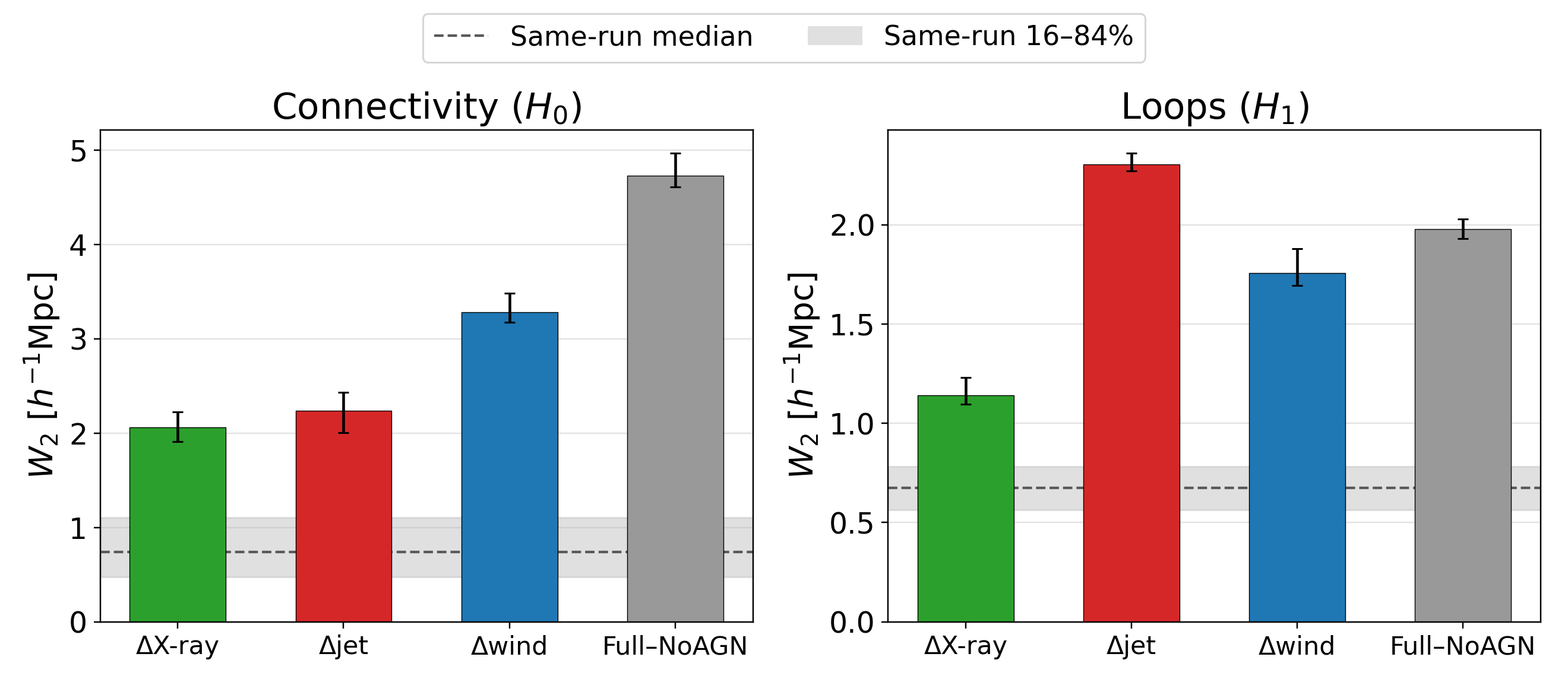}
\end{minipage}
\refstepcounter{figure}\label{fig:fixed_mass_bin_wasserstein_appendix}
\vspace{0.2em}
\begin{minipage}{0.96\textwidth}
\footnotesize
\textbf{FIG.~\thefigure.} Fixed-host-halo-mass-bin decomposition of the 2-Wasserstein distances for the total-galaxy catalog. Each displayed panel follows the same distance convention as the main-text transport figures, including a fourth bar for the direct Full Physics-to-No-AGN comparison, but is restricted to one host-halo-mass bin; the bin labels above the panels give the corresponding $\log_{10}(M_{\mathrm{h}}/M_{\odot})$ ranges. The bin-resolved transport costs provide a quantitative counterpart to the fixed-host-mass Betti curves in Fig.~\ref{fig:exp4_betti_fixed_mh}, showing that the mass-dependent channel hierarchy is visible in the $H_1$ sector: in the lower-mass bins, the wind step provides the largest transport contribution, while in bin 4 the jet step exceeds the wind contribution. This reordering is consistent with efficient AGN kinetic feedback becoming common in sufficiently massive systems, and therefore contributing most strongly to the topological response in the highest-mass halos. The direct comparison exceeds the same-run reference in every bin and dimension shown.
\end{minipage}
\vspace{-0.2em}
\end{widetext}

\FloatBarrier

\section{Central--Satellite Composition and Radial Diagnostics}\label{app:composition}
The fixed-number-density construction used for Fig.~\ref{fig:quenched_absolute} controls the number of tracers entering each topological comparison, but it does not fix the central--satellite composition of the rank-selected star-forming or quenched samples. Table~\ref{tab:sat_fraction} summarizes the corresponding satellite-fraction diagnostics. We define
\begin{equation}
f_{\rm sat}^{\rm Q}
=
\frac{N_{\rm sat}^{\rm Q}}
     {N_{\rm sat}^{\rm Q}+N_{\rm cen}^{\rm Q}},
\qquad
f_{\rm sat}^{\rm SF}
=
\frac{N_{\rm sat}^{\rm SF}}
     {N_{\rm sat}^{\rm SF}+N_{\rm cen}^{\rm SF}}.
\end{equation}
Equivalently, these quantities are $P(\mathrm{satellite}\mid\mathrm{quenched})$ and $P(\mathrm{satellite}\mid\mathrm{star\ forming})$, respectively. They are conditional on different selected samples and therefore are not required to sum to unity. The quenched satellite fraction rises from $0.504$ in Full Physics to $0.894$ in No-AGN, so the Full Physics quenched sample is substantially more central-rich than the corresponding No-AGN sample. This quantity should not be interpreted as a direct central-quenching efficiency; measuring that would require the central-normalized fraction $P(\mathrm{quenched}\mid\mathrm{central})$ or equivalently $N_{\rm Q}^{\rm cen}/N_{\rm all}^{\rm cen}$.

\begin{center}
\begingroup
\refstepcounter{table}\label{tab:sat_fraction}
\begin{minipage}{0.98\columnwidth}
\raggedright
\textbf{TABLE~\thetable.} Central--satellite composition of the fixed-abundance quenched and star-forming catalogs used in the topological analysis. Each realization contains $N_{\rm Q}=1457$ and $N_{\rm SF}=4863$ selected galaxies. The satellite fractions are defined as $f_{\rm sat}^{\rm Q}=N_{\rm sat}^{\rm Q}/(N_{\rm sat}^{\rm Q}+N_{\rm cen}^{\rm Q})$ and $f_{\rm sat}^{\rm SF}=N_{\rm sat}^{\rm SF}/(N_{\rm sat}^{\rm SF}+N_{\rm cen}^{\rm SF})$.
\end{minipage}

\vspace{0.2em}

\resizebox{\columnwidth}{!}{%
\begin{tabular}{lrrrrrr}
\hline
Run
& $N_{\rm sat}^{\rm Q}$
& $N_{\rm cen}^{\rm Q}$
& $f_{\rm sat}^{\rm Q}$
& $N_{\rm sat}^{\rm SF}$
& $N_{\rm cen}^{\rm SF}$
& $f_{\rm sat}^{\rm SF}$ \\
\hline
Full Physics & $735$ & $722$ & $0.504$ & $1155$ & $3708$ & $0.238$ \\
No-Xray & $1005$ & $452$ & $0.690$ & $1203$ & $3660$ & $0.247$ \\
No-Jet & $1231$ & $226$ & $0.845$ & $1214$ & $3649$ & $0.250$ \\
No-AGN & $1302$ & $155$ & $0.894$ & $1257$ & $3606$ & $0.258$ \\
\hline
\end{tabular}%
}
\endgroup
\end{center}

For the radial diagnostic quoted in the main text, the host-halo-mass distributions of quenched satellites are broadly similar across feedback realizations: the median $\log_{10}(M_{200c}/M_{\odot})$ values are $13.36$, $13.52$, $13.62$, and $13.58$ for Full Physics, No-Xray, No-Jet, and No-AGN, respectively. This indicates that the radial trend is not primarily driven by selecting a different host-mass regime. A direct halo-centric radial check supports the same interpretation: the median $R/R_{200c}$ of quenched satellites is $1.061$ in Full Physics and $0.674$ in No-AGN. Thus, even within the satellite component itself, quenched galaxies in Full Physics are less centrally concentrated than in No-AGN.

\FloatBarrier

\section{Reduced-Density Check Against Same-Run Subsampling}\label{app:top50_masscut}
To test whether the topological inference is dominated by finite-sampling fluctuations from the lowest-mass satellite galaxies, and to approximate a stricter stellar-mass completeness cut, we repeat the satellite analysis using only the top $50\%$ of the satellite population ranked by stellar mass in each feedback realization. Under this density cut, the effective minimum stellar-mass threshold increases by nearly an order of magnitude, from approximately $2.6\times10^8\,M_{\odot}$ in the full-density satellite catalog to $2.2\times10^9\,M_{\odot}$. Because halving the tracer density increases the typical inter-galaxy separation and reduces the number of topological features, the same-run finite-sampling reference is expected to broaden.

Figure~\ref{fig:app_top50_masscut} shows that the diagnostic $H_1$ transport signal nevertheless remains above this elevated internal finite-sampling reference. The median transport costs for all three feedback steps exceed the upper edge of the same-run finite-sampling reference. The kinetic-jet step ($\Delta$jet) is the most persistent in this reduced-density catalog: its full 16--84\% interval clears the reference band, whereas the lower bounds for the X-ray and wind steps lie closer to the broadened same-run reference band. This behavior is consistent with the main-text interpretation that the kinetic jet mode is the feedback channel most closely associated with the feedback-dependent spatial difference in massive halos and in the higher-mass satellite-tracer diagnostic.

The corresponding $H_1$ Betti curves show that the qualitative geometric structure is also preserved under the $50\%$ mass cut. The satellite sample retains a visible two-scale response, shifted toward larger filtration scales as expected for a lower-density tracer set. The star-forming and quenched projections within this reduced satellite catalog remain a same-realization subsampling check rather than the basis of the main population inference: the main star-forming/quenched result comes from the total-galaxy catalog in Fig.~\ref{fig:quenched_absolute}. These results indicate that the satellite-tracer diagnostic is not driven solely by the faintest satellites, but remains visible in a higher-mass satellite subsample relevant to stricter observational mass limits.

\section{Methodological Illustration of Inverse Mapping}\label{app:inverse_mapping}
The statistical claims of this work rest entirely on ensemble-level diagnostics---PD/$W_2$ distances, same-run finite-sampling references, fixed-host-mass total-galaxy comparisons, population splits, and $W_2$-matching decompositions---rather than on the identification of individual generators with specific real-space configurations. The local inverse analysis shown in Fig.~\ref{fig:local_inverse_topological_expansion} is included only as a methodological illustration of how such a mapping might be visualized in future work.

Figure~\ref{fig:local_inverse_topological_expansion} illustrates this possibility as a proof of concept. We select one illustrative group-scale halo from the No-AGN realization with $\log_{10}(M_{200c}/M_{\odot})=13.118$ and compare the satellite configuration inside the same fixed real-space aperture in the Full Physics realization. The No-AGN aperture contains 35 satellites and supports a highlighted local $H_1$ loop born at $\alpha_b=0.737\,h^{-1}\mathrm{Mpc}$, while the Full Physics aperture contains 21 satellites and a highlighted local $H_1$ loop born at $\alpha_b=0.959\,h^{-1}\mathrm{Mpc}$. This example illustrates that persistent-homology generators can sometimes be visually associated with local galaxy configurations, but it should not be interpreted as a statistically typical cross-realization generator match.

This local visualization is not a statistical estimator. Local generator matching requires choices about aperture definition, halo centering, and cross-realization association, and individual $H_1$ generators need not have unique one-to-one real-space counterparts across feedback models. The global distribution of birth-scale shifts for high-birth matched generators (right panel of Fig.~\ref{fig:local_inverse_topological_expansion}) has median $\Delta\alpha_b=-0.005\,h^{-1}\mathrm{Mpc}$ and $P(\Delta\alpha_b>0)=0.42$, which does not establish a universal positive birth-scale shift at the ensemble level. A systematic inverse-mapping analysis---aggregating such local information across many halos with controlled aperture and matching definitions---is a possible extension of this work, but lies outside the scope of the present analysis.

\begin{widetext}
\vspace{0.1em}
\centering
\includegraphics[width=0.80\textwidth]{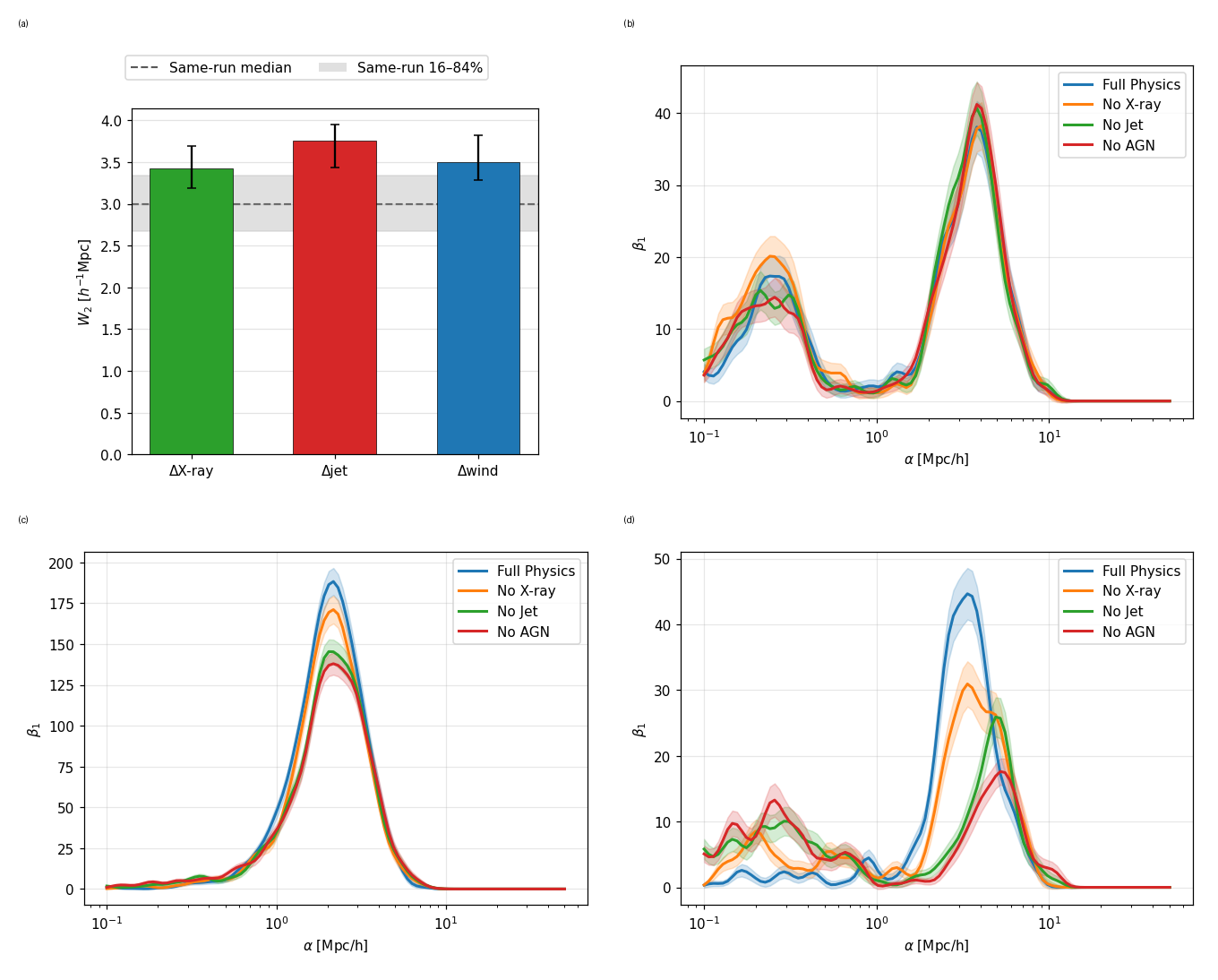}
\refstepcounter{figure}\label{fig:app_top50_masscut}
\vspace{0.15em}
\begin{minipage}{0.94\textwidth}
\scriptsize
\textbf{FIG.~\thefigure.} Same-realization subsampling check using the top $50\%$ of the satellite population ranked by stellar mass ($M_* \gtrsim 2.2\times10^9\,M_{\odot}$). Panel (a) shows the 2-Wasserstein distances for the $50\%$ satellite catalog. Despite the broadened same-run finite-sampling reference produced by the lower tracer density, the median $H_1$ transport costs for all three feedback steps exceed the empirical reference. Panels (b)--(d) show the corresponding $H_1$ Betti curves for the reduced satellite catalog and its star-forming and quenched projections, respectively.
\end{minipage}

\vspace{0.45em}

\includegraphics[width=0.86\textwidth,trim=0 0 0 41.0bp,clip]{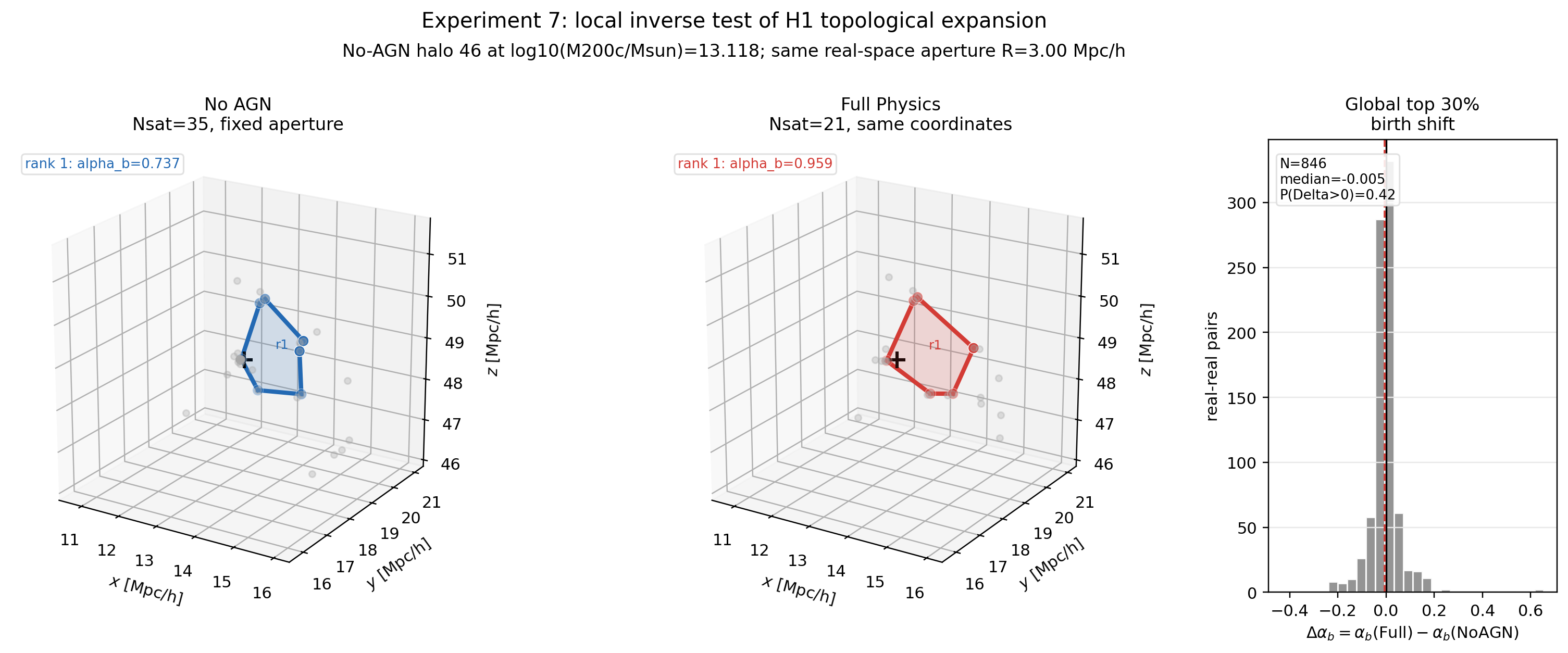}
\refstepcounter{figure}\label{fig:local_inverse_topological_expansion}
\vspace{0.15em}
\begin{minipage}{0.94\textwidth}
\scriptsize
\textbf{FIG.~\thefigure.} Proof-of-concept illustration of TDA inverse mapping for the satellite $H_1$ response. The left and middle panels compare the same real-space aperture around one illustrative No-AGN group-scale halo in the No-AGN and Full Physics realizations. In this aperture, the Full Physics catalog contains fewer satellites (21 versus 35), and a highlighted local $H_1$ loop is born at a larger filtration scale ($0.959$ versus $0.737\,h^{-1}\mathrm{Mpc}$). The right panel shows the distribution of birth-scale shifts for globally matched high-birth generators (median $\Delta\alpha_b=-0.005\,h^{-1}\mathrm{Mpc}$, $P(\Delta\alpha_b>0)=0.42$), emphasizing that this local visual correspondence does not extend to a universal global trend. The left and middle panels are illustrative only; the ensemble-level statistical claims in this paper do not rely on this local inverse-mapping example.
\end{minipage}
\vspace{-0.2em}
\end{widetext}

\FloatBarrier

\section{Same-Tracer Two-Point Correlation Check for Satellites}\label{app:satellite_2pcf}
The two-point correlation function shown in Fig.~\ref{fig:2pcf} is measured on the untrimmed total-galaxy catalog, whereas the satellite topological results in Sec.~\ref{sec:results2} use the fixed-density satellite catalog. Because these are two different tracer populations, the contrast drawn in Sec.~\ref{sec:results2} between the satellite topological response and ``the two-point correlation function'' does not by itself demonstrate that the same pattern holds for a two-point statistic evaluated on the same tracers. Here we repeat the two-point correlation measurement directly on the fixed-density satellite catalog used for the satellite TDA analysis, using the identical natural estimator and same repeated-subsampling error estimate as Fig.~\ref{fig:2pcf}.

Figure~\ref{fig:xi_ss_satellite} shows the resulting satellite-satellite correlation function $\xi_{ss}(r)$. At the smallest scales probed ($r\lesssim0.2\,h^{-1}\mathrm{Mpc}$), the weaker-feedback realizations show a clear excess over Full Physics, consistent in direction with the total-galaxy result in Fig.~\ref{fig:2pcf}. At larger scales, including the $\alpha\sim1$--$3\,h^{-1}\mathrm{Mpc}$ range where the satellite Betti-curve tests (Sec.~\ref{sec:results2}) find a significant cumulative Full Physics-to-No-AGN difference, the satellite-only fractional differences are modest ($\sim5$--$20\%$) and are largely consistent with the subsampling-based error bars. Because the fixed-density satellite catalog ($N=2422$) is substantially smaller than the untrimmed total-galaxy catalog used in Fig.~\ref{fig:2pcf} ($N=6922$), this direct same-tracer check has limited statistical power at these scales and can neither confirm nor rule out a comparable two-point signal there.

This limitation is itself informative: on this same, modest-sized satellite sample, the $D_{\max}$/$T_{\rm area}$ functional tests applied to the Betti curves (Sec.~\ref{sec:results2}) reach formal, Holm-corrected significance for the cumulative comparison at $\alpha\sim1$--$3\,h^{-1}\mathrm{Mpc}$, while a direct two-point measurement on the same tracers lacks the statistical power to resolve this scale range. This is consistent with persistence diagrams extracting more information per tracer from the birth-death structure of topological features than a two-point statistic alone, though a larger satellite sample would be needed to determine conclusively whether a two-point signal is present at these scales.

\begin{center}
\includegraphics[width=0.96\columnwidth]{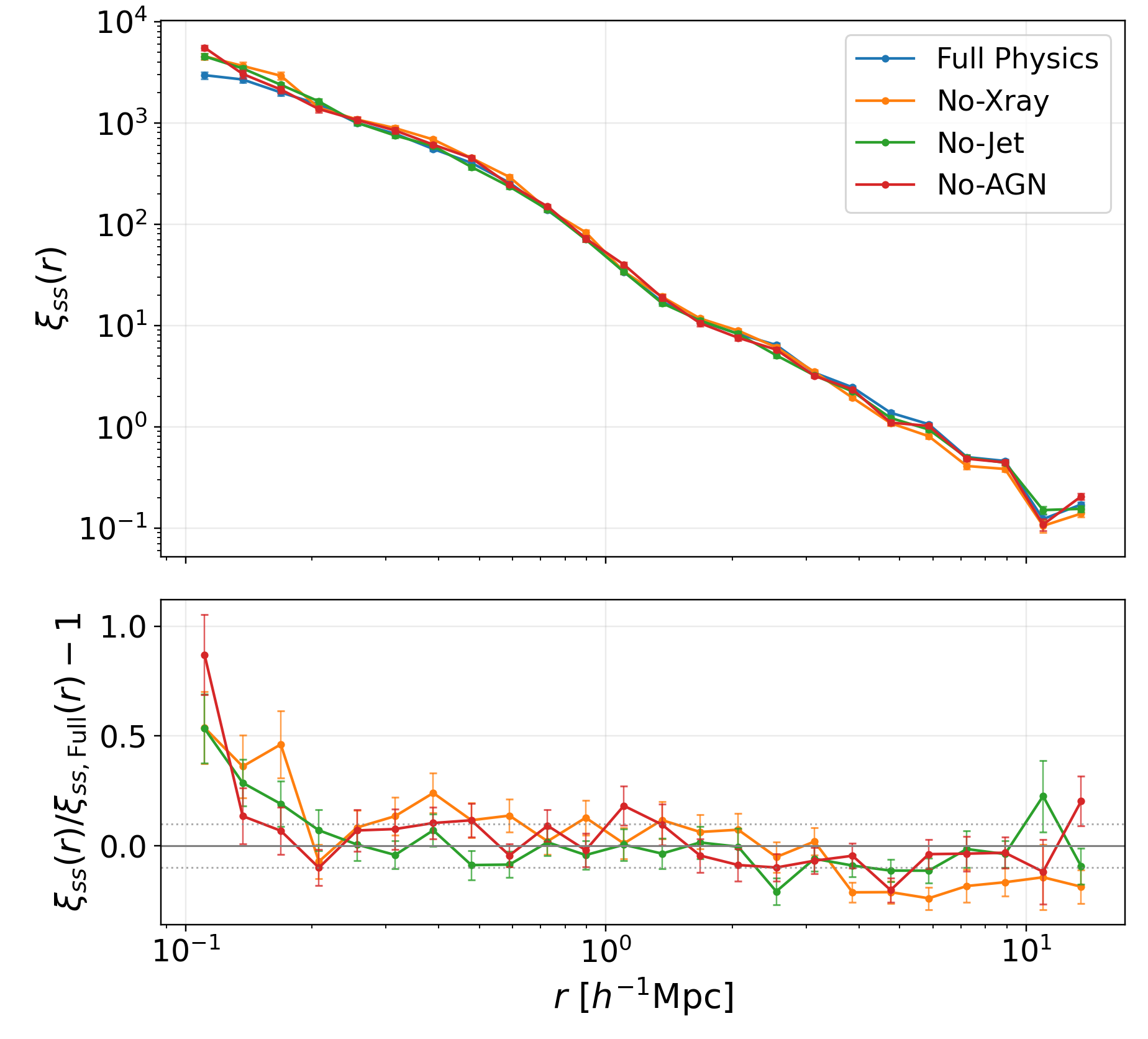}
\refstepcounter{figure}\label{fig:xi_ss_satellite}
\vspace{0.1em}
\begin{minipage}{0.96\columnwidth}
\footnotesize
\textbf{FIG.~\thefigure.} Satellite-satellite two-point correlation function $\xi_{ss}(r)$, measured on the same fixed-density satellite catalog ($N=2422$) used for the satellite TDA analysis in Sec.~\ref{sec:results2}, following the same convention as Fig.~\ref{fig:2pcf}. Error bars show the standard deviation over 50 random $80\%$ subsamples. A clear small-scale excess for the weaker-feedback realizations is visible below $r\sim0.2\,h^{-1}\mathrm{Mpc}$, consistent with Fig.~\ref{fig:2pcf}; at larger scales the fractional differences are modest and largely within the subsampling-based uncertainty, reflecting the smaller tracer count of the satellite catalog relative to the total-galaxy sample.
\end{minipage}
\end{center}

\FloatBarrier
\clearpage

\bibliography{references}

\end{document}